\documentstyle[prb,aps,epsf,multicol]{revtex}
\begin{document}
\title{Spectral functions of the 1D Hubbard model in the 
$U\rightarrow +\infty$ limit: How to use the factorized wave-function}

\author{
  Karlo Penc$^{a}$\cite{*}, Karen Hallberg$^{a}$, Fr\'ed\'eric Mila$^{b}$
  and Hiroyuki Shiba$^{c}$
}

\address{
     $(a)$ Max-Planck-Institut f\"ur Physik komplexer Systeme,
          Bayreuther Str.~40, 01187 Dresden (Germany) \\
     $(b)$ Laboratoire de Physique Quantique, Universit\'e Paul Sabatier,
     31062 Toulouse (France) \\
     $(c)$ Tokyo Institute of Technology, Department of Physics,
           Oh-okayama, Meguro-ku, Tokyo 152 (Japan)
}
\date{December 30, 1996}
\maketitle

\begin{abstract}
We give the details of the calculation of the spectral functions of the 1D
Hubbard model using the spin-charge factorized wave-function
for several versions of the $U\rightarrow +\infty$ limit. 
The spectral functions are expressed as a convolution of charge 
and spin dynamical correlation functions. 
A procedure to evaluate these correlation functions very accurately 
for large systems is developed, and analytical results are 
presented for the low energy region. These results are fully consistent 
with the conformal field theory. We also propose a direct method of 
extracting the exponents from the matrix 
elements in more general cases. 
\end{abstract}


\widetext
\begin{multicols}{2}
\narrowtext

\section{Introduction}

  After the recent photoemission experiments\cite{exp,exp2}
 on quasi one-dimensional
 materials, the need of understanding the dynamical spectral functions of
strongly correlated electron systems has arised. While the low energy 
behavior is usually well described within the framework of the Luttinger 
liquid theory,\cite{voit,schonhammer,voitrev}
the experimentally relevant higher energies ($\approx 100$ meV) can be 
calculated for example by diagonalizing small clusters\cite{ED} or by 
Quantum Monte-Carlo calculations.\cite{montecarlo}
Unfortunately, both methods have limitations either given by the
small size of the system or by statistical errors and use of 
analytic continuation. 
Even for the Bethe-Ansatz solvable models, where the excitation spectra 
can be calculated, the problematic part of calculating the matrix elements
remains: The wave functions are required, and they are 
simply too complicated. There is, however, a special class of models,
where the evaluation of the matrix elements is made possible through a 
relatively simple factorized form of the wave-function, and some results 
were already published by Sorella and Parolla\cite{sorella2} for the 
insulating half-filled case and by the authors\cite{local,shaba} away from
 half-filling.

The dynamical, zero temperature one-particle spectral 
functions can be defined as the imaginary parts of 
the time ordered Green's function:
\[
 \begin{array}{lc}
    A(k,\omega) =  \displaystyle{\frac{1}{\pi}} {\rm Im} G(k,\omega), 
    &\mbox{for $\omega>\mu$};  \\
    &\\
    B(k,\omega) =  -\displaystyle{\frac{1}{\pi}}{\rm Im} G(k,\omega),
    &\mbox{for $\omega<\mu$}. \\
  \end{array}
\]
$A(k,\omega)$ is measured in angular resolved inverse photoemission 
experiments and can be calculated from the Lehmann representation:
\[
 A(k,\omega)  =
 \sum_{f,\sigma} 
  \left| \langle f,N \!+\! 1| a^\dagger_{k,\sigma} |0,N\rangle \right|^2
  \delta(\omega \!-\! E^{N+1}_f \!+\! E^N_0) , 
\]
while $B(k,\omega)$ is measured in the angular resolved photoemission 
experiments and is given by:
\[  
 B(k,\omega) =
 \sum_{f,\sigma} 
  \left| \langle f,N \!-\! 1| 
     a^{\phantom{\dagger}}_{k,\sigma} |0,N\rangle \right|^2
    \delta(\omega \!-\! E^{N}_0 \!+\! E^{N-1}_f) .
\]
Here $N$ is the number of electrons, $f$ denotes the final states and 
$a_{k,\sigma}$ destroys an electron with momentum $k$ and spin $\sigma$. 
If the spectral functions are known, the time ordered Green's 
function can be obtained from
\begin{equation}
  G(k,\omega) = 
  \int_{\mu}^{+\infty} {\rm d} \omega' 
  \frac{A(k,\omega')}{\omega \!-\! \omega' \!+\! i\delta}
+ \int_{-\infty}^{\mu} {\rm d} \omega' 
  \frac{B(k,\omega')}{\omega \!-\! \omega' \!-\! i\delta} .
 \label{eq:GAB}
\end{equation}

  The special models for which the matrix elements can be calculated are:

i) The Hubbard model, defined as usual:
\begin{equation}
  H= -t \sum_{i,\sigma} 
   \left(
    a^\dagger_{i+1,\sigma} a^{\phantom{\dagger}}_{i,\sigma}
   + h.c. \right)
   + U \sum_{i} n_{i,\uparrow} n_{i,\downarrow} ,
\end{equation}
in the limit  $U/t \rightarrow +\infty$;

ii) The anisotropic $t-J$ model 
\begin{eqnarray}
H_{tJ} &=& -
  t \sum_{i, \sigma} 
    (\tilde a_{i,\sigma}^\dagger  
      \tilde a_{i+1,\sigma}^{\vphantom{\dagger}} + {\rm h.c.})
    \nonumber\\
  && +  
    \sum_{i} \sum_{\alpha=x,y,z}
       J^\alpha \left( S_i^\alpha S_{i+1}^\alpha 
     - \case{1}{4} \delta_{\alpha,z} n_i n_{i+1}\right) ,
  \label{eq:Heff}
\end{eqnarray}
in the limit $J^\alpha \rightarrow 0$, where $\tilde a_{i,\sigma}$ are 
the usual projected operators. Actually, the Hubbard model in the 
large $U$ limit can be mapped onto a strong coupling model usually 
identified as the $t-J$ model plus three-site terms using a canonical 
transformation,\cite{harris,oles} where  $J=4 t^2/U$ is small;

iii) An extension of the $t-J$ model first proposed by 
Xiang and d'Ambrumenil,\cite{XA} defined by the Hamiltonian
\begin{eqnarray}
  {\cal H}&=& -t \sum_{i,\sigma} 
   \left(
    \tilde a^\dagger_{i,\sigma} \tilde a^{\phantom{\dagger}}_{i+1,\sigma}
   + h.c. \right) \nonumber\\
   & & + \sum_{i>j} \sum_{\alpha=x,y,z}
     J^\alpha \left( S_i^\alpha S_{i+j}^\alpha 
     - \case{1}{4} \delta_{\alpha,z} n_i n_{i+j}\right) {\cal P}_{i,j} ,
   \label{eq:xiang}
\end{eqnarray}
where  
${\cal P}_{i,j} = \prod_{j'=1}^{j-1} (1 \! - \! n_{i+j'})$
in the exchange part of the Hamiltonian ensures that two spins 
interact as long as there is no other spin between them. 
The motivation to study this model is that, unlike the infinite $U$ 
Hubbard model, there is a finite 
energy $J$ associated with spin fluctuations, and this will give 
us useful indications about the finite $U$ Hubbard model. 

 From the models defined above, the Hubbard model is the most relevant one.
It plays a central role as the generic model of strongly correlated
electron systems. Even though it is comparatively simple, it is very 
difficult
to solve except for the  one dimensional case, where it is solvable by 
Bethe Ansatz.\cite{liebwu} Unfortunately, the Bethe ansatz solution is not 
convenient for direct computation of spectral functions, therefore an 
alternative approach was needed.
In the limit of small $U$ one can use the renormalization 
group\cite{solyom}
to show that the Hubbard model belongs to the universality class of the
Tomonaga-Luttinger model,\cite{tomlut} usually referred to  
as Luttinger-liquid.\cite{haldane} The Luttinger liquids are 
characterized by
power-law decay of correlation functions, and nonexistence of 
quasiparticles.\cite{Greensfunc} The underlying conformal field theory 
can be used to relate the exponents to finite-size corrections of 
the energy and momentum.\cite{woyna89,schulz,frahm,kawakami}
 This gives consistent results not only with the renormalization group in 
the weak coupling regime,\cite{penc93} but also with the special case of
$U/t \rightarrow +\infty$, where the exponents of the static correlations
could be obtained using a factorized wave 
function.\cite{shiba,parola90,ren} 

Actually, the spin-charge factorized wave function also describes the 
excited states as well,\cite{woyna82} and it can be used to calculate
the dynamical spectral functions as well. The spectral functions 
obtained in 
this way are very educative and in some sense, unexpected. 
For example, it turns out that the spectrum contains 
remnants of bands\cite{shaba} crossing the Fermi energy at $3k_F$ - 
the so called shadow bands. Also it gives information on the applicability 
of the 
power-law Luttinger liquid correlation function.\cite{local} 
The aim of this paper is not only to give the details of the 
calculation, that can be useful for other correlation 
functions, but also to present some new results on the low energy behavior
of the charge and spin part (both for the isotropic Heisenberg and XY 
spin model).

 The paper is organized as follows: In Section~\ref{sec:wavefunc} we 
review the factorized wave function and in Section~\ref{sec:specfunc}
we show how the spectral functions can be given as a convolution of 
spin and charge parts. Sections~\ref{sec:charge} and \ref{sec:spin}
are devoted to the detailed analysis of the charge and spin parts.
The relation to the results obtained from the finite-size corrections and
conformal field theory is discussed in Section~\ref{sec:CFT}. Finally,
in Section~\ref{sec:conclu} we present our conclusions.

\section{The factorized wave function}
\label{sec:wavefunc}

It has been shown,\cite{shiba,woyna82} by using the Bethe ansatz solution, 
that the ground state wave function of the Hubbard model in the 
$U \rightarrow +\infty$ limit can be constructed
 as a product of a spinless fermion wave function 
$|\psi\rangle$ and a
squeezed spin wave function $|\chi\rangle$. 
This can be alternatively seen using perturbational arguments\cite{shiba} 
and then extended to the $t-J$ model in the 
 $J\rightarrow 0$ limit.
Moreover, the wave function of the excited
states are also factorized:\cite{sorella2,woyna82}
\begin{equation}
  |N \rangle = |\psi^{N}_{L,Q}(\{ {\cal I} \})\rangle 
         \otimes |\chi^{N_\downarrow}_N(Q,\tilde f_Q) \rangle .
  \label{eq:OSwfLHB}
\end{equation} 

 The spinless fermion wave function
$|\psi\rangle$ describes the charges and is an eigen function of 
$N$ noninteracting spinless fermions on $L$ sites with momenta
\begin{equation}
  k_j L = 2 \pi {\cal I}_j + Q ,
  \label{eq:kIQ}
\end{equation}
where the ${\cal I}_j$  are integer quantum numbers and $j=1,2,\dots N$.
The charge part is not fully decoupled from the spin wave function 
$|\chi\rangle$, as the momentum  
$Q=2 \pi {\cal J}/N$ (${\cal J}=0,1,\dots,N-1$ ) 
of the spin wave function imposes a
twisted boundary condition on the spinless fermion wave-function 
(each fermion
hopping from site $L-1$ to site $0$ will acquire a phase $e^{iQ}$) to 
ensure periodic boundary conditions for the original problem.
The energy of the charge part is 
\begin{equation}
  E^N_c = -2 t \sum_{j=1}^{N} \cos k_j ,
  \label{eq:ELHB}
\end{equation}
and the momentum reads $P^N_c = \sum_{j=1}^{N} k_j$, or, 
using Eq.~(\ref{eq:kIQ}):
\begin{equation}
  P^N_c = \frac{2 \pi}{L} \sum_{j=1}^{N} {\cal I}_j + \frac{N}{L} Q 
    = \frac{2 \pi}{L} \left( \sum_{j=1}^{N} {\cal I}_j + {\cal J} \right) .
  \label{eq:PLHB}
\end{equation}

On the other hand, the spin wave functions $|\chi\rangle$ are characterized
by the number of down spins $N_\downarrow$, the total momentum $Q$, and 
the quantum number $\tilde f_Q$ within the subspace of momentum $Q$. They
are eigenfunctions of the Heisenberg Hamiltonian
\begin{equation}
H_s =  \sum_{i=1}^N \sum_{\alpha=x,y,z}
       \tilde J^\alpha \left( S_i^\alpha S_{i+1}^\alpha 
     - \case{1}{4} \delta_{\alpha,z} \right) .
  \label{eq:Hseff}
\end{equation}
with eigenenergies $E_s$.
$\tilde J^\alpha$ depends on the actual charge wave 
function $|\psi\rangle$. 
In the case of the $U\rightarrow +\infty$ Hubbard model,
\begin{equation}
  \tilde J = \frac{2t^2}{U} \frac{1}{N} \sum_{i,\delta=\pm 1} 
      \langle \psi |
         n_i n_{i+\delta} 
        - b^{\dagger}_{i+\delta}
          n_{i}
          b^{\phantom{\dagger}}_{i-\delta} 
      |\psi\rangle ,
  \label{eq:JeffH}
\end{equation}
where $b^\dagger_j$ and $b^{\phantom{\dagger}}_j$ are the operators of
spinless fermions at site $j$.
For the ground state $|\psi^{\rm GS}\rangle $ it reads 
$\tilde J^\alpha = n (4t^2/U) [1-\sin(2 \pi n)/(2 \pi n)]$, 
where $n=N/L$ is the density.

For the $t-J$ model:
\begin{equation}
  \tilde J^\alpha = J^\alpha 
  \sum_{i} \langle \psi | n_i n_{i+j} |\psi\rangle ,
  \label{eq:JefftJ}
\end{equation}
and for the ground state 
$\tilde J^\alpha = J^\alpha n [1-\sin^2(\pi n)/(\pi n)^2]$.
For the model of Xiang and d'Ambrumenil $\tilde J^\alpha = n J^\alpha $ 
and is independent of the charge part. 
 The energy of the factorized wave function is then given as the 
sum of the charge and spin energies, with the assumption that 
the correct $\tilde J$ is 
chosen. If $U \rightarrow +\infty$ or $J \rightarrow 0$, then the spectrum 
collapses and we can assume all the spin states degenerate, simplifying 
considerably some of the calculations to be presented later.

Furthermore, we choose $N$ to be of the form $4l + 2$ ($l$ integer), when 
the ground state is unique. 
Then in the ground-state 
the spinless fermion wave-function $|\psi_{L}^{N,\rm GS}\rangle$ is
described by the quantum numbers $Q=\pi$ and 
$\{{\cal I} \}=\{-N/2,\dots,N/2-2,N/2-1 \}$, so that the distribution 
of the $k_j$'s is
symmetric around the origin and we choose the spin part as the 
ground-state of the
Heisenberg model according to Ogata and Shiba's prescription.\cite{shiba}
 This choice of the spin wave function makes the difference between the 
$U \rightarrow +\infty$ and $U = +\infty$ (the so called $t$ model) limits.

  The price we have to pay for such a simple wave function is that the
representation of real fermion operators $a^\dagger_{j,\sigma}$ in the 
new basis becomes complicated. As a first step, we can write 
$a^\dagger_{j,\sigma}$ as 
$a^\dagger_{j,\sigma} = a^\dagger_{j,\sigma} (1-n_{j,\bar\sigma}) + 
  a^\dagger_{j,\sigma} n_{j,\bar\sigma}$, where 
$a^\dagger_{j,\sigma} (1-n_{j,\bar\sigma}) $ creates a fermion at
an unoccupied site and the $a^\dagger_{j,\sigma} n_{j,\bar\sigma}$ adds a 
fermion at an already occupied site,
thus creating a doubly occupied site. $\bar\sigma$ means the spin state
opposite to $\sigma$. This latter process  gives contributions
to the spectral functions in the upper 
Hubbard band, $A^{\rm UHB}(k,\omega)$ which can be calculated in a 
similar way, but we will not address this issue in the present paper. 

Next, we define the operators $\hat Z^\dagger_{i,\sigma}$ and 
$\hat Z^{\phantom{\dagger}}_{i,\sigma}$ acting on the spin part of the wave
function:
The $\hat Z^{\dagger}_{i,\sigma}$ adds 
a spin $\sigma$ to the beginning of the spin wave function 
$|\chi_N \rangle$ if
$i=0$, or inserts a spin $\sigma$ after skipping the first $i$ spins,  
and makes it $N+1$ long, e.g.:
$\hat Z^\dagger_{0,\sigma}|\uparrow\downarrow \rangle =
  |\sigma\uparrow\downarrow \rangle $
and 
 $\hat Z^\dagger_{1,\sigma} |\uparrow\downarrow \rangle =
  |\uparrow\sigma\downarrow \rangle $. 
The $\hat Z_{i,\sigma}$ is 
  defined as the adjoint operator of 
   $\hat Z^\dagger_{i,\sigma}$, i.e. it removes a spin from site $i$.

Then, to create a fermion at the empty site $j=0$, we need to create
 one spinless fermion
with operator $b^\dagger_0$ and to add a spin $\sigma$ to 
the spin wave function
with operator $\hat Z^\dagger_{0,\sigma}$:
\begin{equation}
  a^\dagger_{0,\sigma} (1-n_{0,\bar\sigma})
      =  \hat Z^\dagger_{0,\sigma} b^\dagger_0 .
    \label{eq:adecLHB}
\end{equation}
The apparent simplicity is lost for  $a^\dagger_{1,\sigma}$. 
Then, apart from creating a spinless fermions with $b^\dagger_1$ in 
the charge part, we have to consider the following two possibilities: 
either the $j=0$ site is empty, and
with $a^\dagger_{1,\sigma}$ we create a spin at the beginning of the 
spin wave function
with $\hat Z^\dagger_{0,\sigma}$;
or it is occupied, and we insert a spin between the first and second spin
in $|\chi\rangle$ with $\hat Z^\dagger_{1,\sigma}$.  So we end up with
\[
  a^\dagger_{1,\sigma} (1-n_{1,\bar\sigma})
      =  \left[
                 (1-n_0)\hat Z^\dagger_{0,\sigma}  +
                 n_0  \hat Z^\dagger_{1,\sigma}  
                \right] b^\dagger_1 .
\]
Obviously we choose the
$j=0$ in further calculations
for its simplicity. However, one can show that the final
result does not depend on this special choice and the 
translational invariance is preserved even for these complicated operators.

\section{Spectral Functions}
\label{sec:specfunc}

 To use the factorized wave functions in the calculation of the spectral
 function it is more convenient to transfer the $k$ dependence from 
 the $a^\dagger_{k,\sigma}$ operator to the final state: 
\begin{eqnarray}
  A(k,\omega) & =& 
 \sum_{f,\sigma} 
  L \left| \langle f,N\!+\!1| a^\dagger_{0,\sigma} |0,N\rangle \right|^2 
  \nonumber\\
  && \times \delta(\omega \!-\! E^{N+1}_f \!+\! E^N_0)
            \delta_{k, P^{N+1}_f  \!-\!  P_0^N} 
  \nonumber
\end{eqnarray}
and 
\begin{eqnarray}
  B(k,\omega)& =& 
 \sum_{f,\sigma} L
  \left| \langle f,N\!-\!1| a^{\phantom{\dagger}}_{0,\sigma} 
  |0,N\rangle \right|^2
  \nonumber\\
  && \times    \delta(\omega \!-\! E^{N}_0 \!+\! E^{N-1}_f)
     \delta_{k, P_0^N \!-\! P^{N-1}_f} ,
  \nonumber
\end{eqnarray}
 where the  momenta of the final states are $P^{N\pm 1}_f$.

As we already pointed out, the addition of an electron to the ground state
can result in a final state with or without a doubly occupied state. 
Correspondingly, the spectral function has contributions from the 
upper and lower Hubbard bands: 
$A(k,\omega) = A^{\rm UHB}(k,\omega) + A^{\rm LHB}(k,\omega)$.
We will now consider  $A^{\rm LHB}(k,\omega)$ only. 
 From Eqs.~(\ref{eq:OSwfLHB}) and (\ref{eq:adecLHB}) we get the
following convolution as a consequence of the wave function factorization:
\begin{equation}
  A^{\rm LHB}(k,\omega) 
  =
 \sum_{Q,\omega',\sigma} 
  C_{\sigma}(Q,\omega')
  A_{Q}(k,\omega-\omega') ,
  \label{eq:alhbca} 
\end{equation}
and similarly for $B(k,\omega)$:
\begin{equation}
  B(k,\omega) 
  =
 \sum_{Q,\omega',\sigma} 
  D_{\sigma}(Q,\omega')
  B_{Q}(k,\omega-\omega') .
  \label{eq:blhbca} 
\end{equation}
  $A_{Q}(k,\omega)$ and $B_{Q}(k,\omega)$ depend on the spinless fermion 
wave function only:
\begin{eqnarray}
  A_{Q}(k,\omega) 
 & = & L
 \sum_{\{I\}} 
  \left| 
    \langle \psi_{L,Q}^{N+1}(\{I\}) |
    b^\dagger_{0} 
    | \psi_{L,\pi}^{N,{\rm GS}} \rangle
    \right|^2 \nonumber\\
   &&
   \times
  \delta(\omega - E^{N+1}_{f,c} + E^N_{{\rm GS},c}) 
            \delta_{k, P^{N+1}_{f,c}  \!-\!  P^N_{{\rm GS},c}} ,
  \nonumber\\
  B_{Q}(k,\omega) 
 & = & L
 \sum_{\{I\}} 
  \left| 
    \langle \psi_{L,Q}^{N-1}(\{I\}) |
    b^{\phantom{\dagger}}_{0} 
    | \psi_{L,\pi}^{N,{\rm GS}} \rangle
    \right|^2\nonumber\\
   &&\times 
  \delta(\omega - E^N_{{\rm GS},c} + E^{N-1}_{f,c} ) 
  \delta_{k, P^N_{{\rm GS},c} \!-\! P^{N-1}_{f,c}} ,
 \label{eq:defaomegaQ}
\end{eqnarray}
and they are discussed in more detail in the next section 
(Sec. \ref{sec:charge}).

On the other hand, $C_{\sigma}(Q,\omega)$ and $D_{\sigma}(Q,\omega)$ 
are determined by the spin wave function only:
\begin{eqnarray}
  C_{\sigma}(Q,\omega) &=& \sum_{\tilde f_Q}
    \left| \langle \chi_{N+1} (Q,\tilde f_Q) | 
      \hat Z^\dagger_{0,\sigma} 
      | \chi_N^{\rm GS} \rangle
    \right|^2 
  \nonumber \\
   &&\times \delta(\omega - E^{N+1}_{f,s} + E^{N}_{{\rm GS},s} ) , 
  \nonumber\\
  D_{\sigma}(Q,\omega) &=& \sum_{\tilde f_Q}
    \left| \langle \chi_{N-1}(Q,\tilde f_Q) | 
      \hat Z_{0,\sigma} 
      | \chi_{N}^{\rm GS} \rangle
    \right|^2
  \nonumber \\
  &&\times \delta(\omega - E^{N}_{{\rm GS},s} + E^{N-1}_{f,s} ) ,
   \label{eq:defcomegaQ}
\end{eqnarray}
and are analyzed in Sec.~\ref{sec:spin}. Although we do not 
present it here,
a similar analysis can be made for $A^{\rm UHB}(k,\omega)$.

In Eqs.~(\ref{eq:alhbca}) and (\ref{eq:blhbca})
 the simple addition of the spin and charge energies is assumed.
Strictly speaking, this is only valid for the 
$U\rightarrow +\infty$, $J\rightarrow 0$ and the model of  
Xiang and d'Ambrumenil for any $J$. In the other cases the 
dependence of $\tilde J$ on the charge wave function should be
explicitly taken into account. Still, it is a reasonable approximation, 
as the important matrix elements will come from exciting a few
 particle-hole excitations only, which will give finite-size corrections
  to $\tilde J$ in the thermodynamic limit. Furthermore, we are 
  neglecting
the $t^2/U$ corrections to the effective operators\cite{oles} and 
to the wave functions.

The momentum distribution function, 
$n_k = \langle a^{\dagger}_k a^{\phantom{\dagger}}_k \rangle$
can be calculated from the spectral function as
 $ n_k =  \int B(k,\omega) d\omega$,
leading  to a similar expression as used by Pruschke 
and Shiba:\cite{XYtJ} 
\begin{equation}
  n_k =  \sum_Q B_Q(k) D(Q) ,
  \label{eq:nkf}
\end{equation}
where $B_Q(k)=\int B_Q(k,\omega) d \omega$ and similarly 
$D(Q)=\int D(Q,\omega) d \omega$.

The local spectral function $A(\omega) = \frac{1}{L} \sum_{k} A(k,\omega)$
is given by
\begin{equation}
  A(\omega) =   \sum_{Q,\omega',\sigma} C_{\sigma}(Q,\omega')
                A_{Q}(\omega-\omega') ,
  \label{eq:alocal} 
\end{equation}
where $A_{Q}(\omega) = \frac{1}{L} \sum_{k} A_{Q}(k,\omega)$. 
Similar equation holds for $B(\omega)$.

\section{About $A_{Q}(k,\omega)$ and $B_{Q}(k,\omega)$}
\label{sec:charge}

To calculate $A_{Q}(k,\omega)$ and $B_{Q}(k,\omega)$ defined in
Eq.~(\ref{eq:defaomegaQ}), we need to evaluate matrix elements like
$\langle \psi_{L,Q}^{N+1}(\{I\}) | b^\dagger_{0}| \psi_{L,Q'}^{N,\rm GS} 
 \rangle $,
where the two states have different boundary conditions. 
In the ground state $Q'=\pi$, but we will not specify $Q'$ yet.  
To calculate these matrix elements, we need the following anti-commutation 
relation:
  \begin{eqnarray}
    \bigl\{ b^\dagger_{k'}, b^{\phantom{\dagger}}_{k} \bigr\}
    &=& \frac{1}{L}\sum_{j,j'} e^{ik'j'-ikj}  
    \bigl\{ b^\dagger_{j'}, b^{\phantom{\dagger}}_{j} \bigr\} \nonumber\cr
    &=& \frac{1}{L} e^{-i(k'-k)/2}e^{i(Q'-Q)/2}
       \frac{\sin([Q'-Q]/2)}{\sin([k'-k]/2)} .
  \end{eqnarray}
where $k$ and $k'$ are wave-vectors with phase shift $Q/L$ and $Q'/L$, 
respectively, see Eq.~(\ref{eq:kIQ}). For $Q\rightarrow Q'$ the 
anti-commutation relation is the usual one: 
$\bigl\{ b^\dagger_{k'}, b^{\phantom{\dagger}}_{k} \bigr\} 
= \delta_{k,k'}$,
 while for $Q \neq Q'$  the overall phase shift $(Q-Q')/L$ due 
to momentum transfer $Q-Q'$ to the spin degrees of freedom
gives rise to the Anderson's orthogonality catastrophe.\cite{ortho1} 
Then a typical overlap
$    \langle 0|
      b^{\phantom{\dagger}}_{k_N} 
      \dots 
      b^{\phantom{\dagger}}_{k_2}
      b^{\phantom{\dagger}}_{k_1}
      b^\dagger_{k'_1} 
      b^\dagger_{k'_2} 
      \dots 
      b^\dagger_{k'_N}
   |0\rangle 
$, where $|0\rangle$ is the vacuum state, is given by the 
following determinant: 
\[
   \left| 
     \begin{array}{cccc}
       \bigl\{ b_{k'_1,k_1}\bigr\} & 
       \bigl\{ b_{k'_1,k_2}\bigr\} & \dots & 
       \bigl\{ b_{k'_1,k_N}\bigr\} \cr
       \bigl\{ b_{k'_2,k_1}\bigr\} & 
       \bigl\{ b_{k'_2,k_2}\bigr\} & \dots &
       \bigl\{ b_{k'_2,k_N}\bigr\} \cr
        \vdots     &  \vdots     &       & \vdots      \cr
       \bigl\{ b_{k'_N,k_1}\bigr\} &
       \bigl\{ b_{k'_N,k_2}\bigr\} & \dots &
       \bigl\{ b_{k'_N,k_N}\bigr\} 
     \end{array}
   \right| .
\]
Replacing the anticommutator, the determinant above becomes
\begin{eqnarray}
 && L^{-N} 
  e^{i(Q'-Q)N/2}  
  \prod_j e^{-i (k'_j-k_j)/2}
  \sin^N\frac{Q'-Q}{2} 
  \nonumber\\
  && \times
  \left| \begin{array}{cccc}
      \sin^{-1} \frac{k'_1-k_1}{2} &
      \sin^{-1} \frac{k'_1-k_2}{2} & 
      \dots &
      \sin^{-1} \frac{k'_1-k_N}{2}
    \cr
       & & &      
    \cr
      \sin^{-1} \frac{k'_2-k_1}{2} &
      \sin^{-1} \frac{k'_2-k_2}{2} &
      \dots & 
      \sin^{-1} \frac{k'_2-k_N}{2} 
    \cr
       & & &      
    \cr
      \vdots     &  \vdots     &       & \vdots      
    \cr
       & & &      
    \cr
      \sin^{-1} \frac{k'_N-k_1}{2} &
      \sin^{-1} \frac{k'_N-k_2}{2} &
      \dots & 
      \sin^{-1} \frac{k'_N-k_N}{2} 
   \end{array}
   \right| .
 \nonumber
\end{eqnarray}
This determinant is very similar to the Cauchy determinant
 (there  the elements are  $1/(k-k')$ instead of 
$1/\sin(k-k')$) and it can be expressed as a 
product,\cite{ortho} so for the overlap we get:
\begin{eqnarray}
  &\pm& 
  L^{-N} 
  e^{i(Q'-Q)N/2} 
  \sin^N\frac{Q'-Q}{2} 
  \prod_j e^{-i (k'_j-k_j)/2}
  \nonumber \cr
  &\times&
   \prod_{j>i} \sin \frac{k_j-k_i}{2} 
   \prod_{j>i} \sin \frac{k'_j-k'_i}{2} 
   \prod_{i,j} \sin^{-1} \frac{k'_i-k_j}{2} ,
\end{eqnarray}
 where the sign $+$ is for $N=1,4,5,8,9,..$ and $-$ for $N=2,3,6,7,..$.

  Now we turn back to the $A_Q(k,\omega)$. The matrix elements in 
Eq.~(\ref{eq:defaomegaQ}) are

\end{multicols}
\vspace{-0.6truecm}
\noindent\makebox[8.8truecm]{\hrulefill}
\widetext
\begin{eqnarray}
  L \bigl|
   \langle 
     \psi^{N+1}_{L,Q}(\{{\cal I}\}) | b^\dagger_{0}
     | \psi_{L,Q'}^{N ,\rm GS} 
   \rangle
  \bigr|^2 &=& 
  \Bigl|
   \sum_{q'}
   \langle 
     \psi^{N+1}_{L,Q}(\{{\cal I}\}) | b^\dagger_{q'}
     | \psi^{N,\rm GS}_{L,Q'} 
   \rangle
  \Bigr|^2 \nonumber\\
  &=&
  L^{-2N} 
  \sin^{2N}\frac{Q'-Q}{2} 
   \prod_{j>i} \sin^2 \frac{k_j-k_i}{2} 
   \prod_{j>i} \sin^2 \frac{k'_j-k'_i}{2} 
   \prod_{i,j} \sin^{-2} \frac{k'_i-k_j}{2} ,
  \label{eq:aqmatele} 
\end{eqnarray}
\vspace{-0.4truecm}
\hspace*{\fill}\makebox[8.8truecm]{\hrulefill}
\begin{multicols}{2}
\narrowtext\noindent
where $q'$ is a wave vector with phase shift  $Q'/L$. Here we have 
used that 
\begin{eqnarray}
 && \Biggl|
   \sum_{q'}
     e^{iq'/2} 
    \prod_{i'=1}^{N} \sin \frac{k'_{i'} \! - \! q'}{2} 
    \prod_{i=1}^{N+1} \sin^{-1} \frac{k_i \! - \! q'}{2} 
  \Biggr|^2 
   \nonumber\\
   &&
   = 
   L^2 \sin^{-2} \frac{Q' \! - \! Q}{2}  \nonumber 
\end{eqnarray}
holds, independently of the actual quantum numbers $\{{\cal I}\}$ and 
$\{{\cal I'}\}$.

Similarly, for the matrix elements in $B_Q(k,\omega)$ we get:
\begin{eqnarray}
  &&L \left| 
    \langle \psi^{N-1}_{L,Q}(\{I\}) |
    b^{\phantom{\dagger}}_{0} 
    | \psi_{L,Q'}^{N,\rm GS} \rangle
  \right|^2 
  = L^{-2N+2} 
  \sin^{2N-2}\frac{Q' \!-\! Q}{2}
  \nonumber\\
  &&\times 
   \prod_{j>i} \sin^2 \frac{k_j \!-\! k_i}{2} 
   \prod_{j>i} \sin^2 \frac{k'_j \!-\! k'_i}{2} 
   \prod_{i,j} \sin^{-2} \frac{k'_i \!-\! k_j}{2}  .
 \nonumber
\end{eqnarray}

  We are now ready to calculate the spectral functions numerically. 
One has to generate  the quantum numbers ${\cal I}_j$, and evaluate 
the energy, momentum and the expressions above.

 From now on, we will consider $Q'=\pi$.

First of all, it turns out that
 the following sum rules are satisfied for every $Q$:
\begin{eqnarray}
  \int_{-\pi}^{\pi} \frac{dk}{2 \pi} 
  \int_{-\infty}^\infty  d\omega A_{Q}(k,\omega) &=& 1-n ,
  \nonumber\\
  \int_{-\pi}^{\pi} \frac{dk}{2 \pi}
  \int_{-\infty}^\infty  d\omega B_{Q}(k,\omega) &=& n .
  \label{eq:srAB}
\end{eqnarray}

 In the absence of the Anderson orthogonality catastrophe, when $Q=Q'=\pi$,
the contribution to the spectral functions comes from one particle-hole 
excitations only, and  the spectral functions are  
nothing but the familiar
 $\delta(\omega+2 t \cos k)$. 
This is not true any more when we consider $Q \ne \pi$. In that case we 
get contributions from
many particle-hole excitations as well. The largest weight comes 
from the one particle-hole
excitations, and increasing the number of excited holes, the additional 
weight  
decreases rapidly. Although from Eq.~(\ref{eq:aqmatele}) we can calculate 
the matrix elements numerically for all the excitations of the final 
state, its application is
limited to small system sizes (typically $L<30$). It is due to the fact 
that the
time required to generate all the possible states 
(quantum numbers ${\cal I}$) is growing
exponentially. Therefore, in some of the calculations we take into account 
up to three particle-hole excitations only. In Table.~\ref{tab:sr} we give 
the total sum rule
for small sizes in a calculation where we took into account up to
one, two and three particle-hole excitations. We can see that 
the missing weight is really small in the approximation that includes 
up to three particle-hole excitations in the final state. 
So, if we restrict ourselves to a finite number of particle-hole
excitations and introduce the function
\begin{eqnarray}
  g({\cal I}) &=& 
    \prod_{{\cal I'}=-N/2 \atop {{\cal I'}\ne {\cal I}}}^{N/2}
    \sin^{2} \frac{\pi}{L} ({\cal I} \!-\! {\cal I'}) 
   \nonumber\\
   && \times
   \prod_{{\cal I''}=-N/2}^{N/2-1} 
    \sin^{-2}\left([{\cal I} \!-\! {\cal I''}]
                    \frac{\pi}{L} + \frac{Q \! - \! \pi}{2L} 
             \right) ,
  \label{eq:defg}
\end{eqnarray} 
the calculation of the spectral weight becomes simple. The weight of 
the peak corresponding to a one particle-hole excitation can be given as:
\begin{equation}
  A_Q({\cal I}^p,{\cal I}^h) = 
    \frac{ g({\cal I}^p) }{g^({\cal I}^h)} 
    \frac{1}{\sin^2([{\cal I}^h-{\cal I}^p]\frac{\pi}{L})} A_Q^{(0,0)} ,
  \label{eq:def1ph}
\end{equation}
where we have removed the quantum number ${\cal I}^h$ (hole) from
and added ${\cal I}^p$ (particle) to the set $\{{\cal I}\}$  
of the ground-state of $N\pm 1$ fermions, so that the momentum of the 
final state is 
$P^{N+1}_f=k^p- k^h + P_{\rm GS}^{N+1}$ and the energy is
$E^{N+1}_f=E^{N+1}_{\rm GS} - 2t \cos k^p + 2t \cos k^h $, where
the  $P_{\rm GS}^{N+1}=(N+1)Q/L$ is the momentum of the ground state.
Furthermore, $A_Q^{(0,0)}$
is the overlap between the $N$ electron ground state with boundary 
condition $\pi$ and the $N+1$ electron ground state with boundary
 condition $Q$, and will be discussed later.

  Similarly, for the two particle-hole excitations we get:
\end{multicols}
\widetext
\vspace{-0.6truecm}
\noindent\makebox[8.8truecm]{\hrulefill}
\begin{equation}
    A_Q({\cal I}^p_{1},{\cal I}^p_{2},{\cal I}^h_{1},{\cal I}^h_{2}) = 
      \frac{ g({\cal I}^p_{1}) g({\cal I}^p_{2}) }
           { g({\cal I}^h_{1}) g({\cal I}^h_{2}) }
       \frac{ \sin^2([{\cal I}^h_{1} \!-\! {\cal I}^h_{2}]\frac{\pi}{L}) 
              \sin^2([{\cal I}^p_{1} \!-\! {\cal I}^p_{2}]\frac{\pi}{L})} 
            { \sin^2([{\cal I}^p_{1} \!-\! {\cal I}^h_{1}]\frac{\pi}{L}) 
              \sin^2([{\cal I}^p_{1} \!-\! {\cal I}^h_{2}]\frac{\pi}{L})
              \sin^2([{\cal I}^p_{2} \!-\! {\cal I}^h_{1}]\frac{\pi}{L})
              \sin^2([{\cal I}^p_{2} \!-\! {\cal I}^h_{2}]\frac{\pi}{L})} 
      A_Q^{(0,0)}
  \label{eq:def2ph}
\end{equation}
\vspace{-0.4truecm}
\hspace*{\fill}\makebox[8.8truecm]{\hrulefill}

\begin{multicols}{2}
\narrowtext\noindent
with energy and momentum 
\begin{eqnarray}
  E^{N+1} &=&  E^{N+1}_{\rm GS} - 2t 
   \left( \cos k^p_{1}+\cos k^p_{2}-\cos k^h_{1}-\cos k^h_{2} \right) 
     ,
  \nonumber\\
  P^{N+1} &=& k^p_{1}+k^p_{2}-k^h_{1}-k^h_{2} + P^{N+1}_{\rm GS} .
  \nonumber
\end{eqnarray}
The corresponding equations for three or more particle-hole
excitations are similar to those above, but since they are long, we do
not give them here.

\begin{figure}
\epsfxsize=8.5 truecm 
\centerline{\epsffile{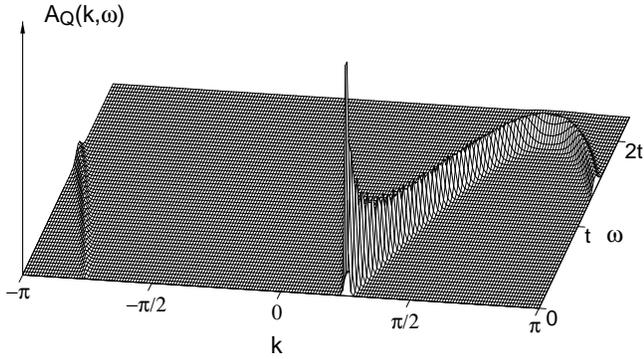}}
\caption{
 $A_Q(k,\omega)$ for $Q=48 \pi/97$ ($\approx\pi/2$), $N=96$ electrons on
$L=192$ sites. We can see the power-law singularity at $k=\pi/4$ and 
that the weight is accumulated along a cosine-like band like structure. 
}
\label{fig:AomkQ}
\end{figure}

\begin{figure}
\epsfxsize=8.5 truecm 
\centerline{\epsffile{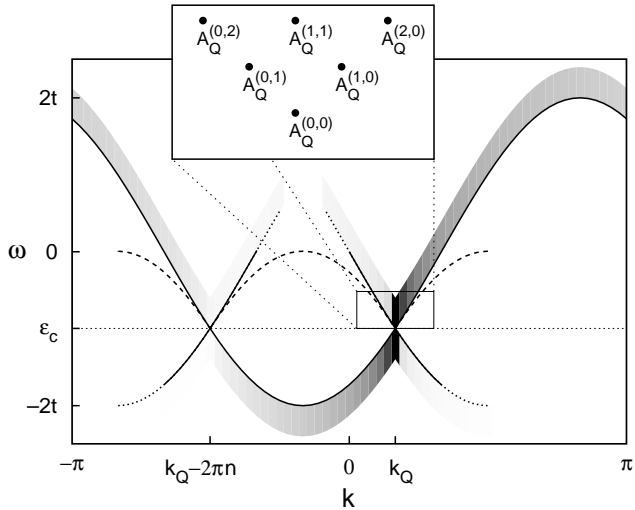}}
\caption{
  Schematic plot of the support of $A_{Q}(k,\omega)$ 
(above $\varepsilon_{c}$) 
and $B_{Q}(-k,\omega)$ (below $\varepsilon_c$)
for $N/L=1/3$ and $Q=\pi/2$. The dominant tower ($p=0$) at $k=k_Q$  and the
sub-dominant tower ($p=-1$) at $k=k_Q-2\pi n$ are shown. The weight mostly
follows the solid lines, and the shadowing represent the intensity. 
Although there are excitations above the dashed line for $A_Q(k,\omega)$ 
as well, the weight associated with them is negligible. The low energy 
part of $A_Q(k,\omega)$ near $k=k_Q$ is enlarged on the insert, where the 
discrete states in the tower of excitations are shown.
}
\label{fig:support}
\end{figure}

  A typical plot of $A_Q(k,\omega)$ is shown in Fig.~\ref{fig:AomkQ}. 
We choose $Q=\pi/2$, which is halfway between the symmetric $Q=0$ and the 
trivial $Q=\pi$ case. In the figure we can  see the singularity near 
the Fermi energy, furthermore the weights are distributed on a cosine-
like band. To make it more clear, in Fig.~\ref{fig:support} we show 
the support of $A_Q(k,\omega)$ and the distribution of the weights.

\subsection{ The weight of the lowest peak} 

 Now, what can we say about $A_Q^{(0,0)}$, the weight of the lowest peak? 
In the ground state the quantum numbers ${\cal I}_j$ and ${\cal I}_j'$ 
are densely packed, and from Eq.~(\ref{eq:aqmatele}) we get 
\begin{eqnarray}
 A_Q^{(0,0)} &=& \frac{ \cos^{2N}(Q/2) }{L^{2N}}  
      \prod_{j=1}^N  \left[\sin^2 \frac{\pi j}{L}\right]^{2 N + 1-2 j}
  \nonumber\\
 && \times \prod_{j=1}^N   
            \left[ 
               \sin^2 \frac{(2 j\!-\!1)\pi\! +\!Q}{2 L} 
               \sin^2 \frac{(2 j\!-\!1)\pi \!-\!Q}{2 L} \right]^{j-N-1}
  \nonumber
\end{eqnarray}
 
 From this we can conclude that $A_Q^{(0,0)}$ is an even function 
of $Q$ and  $A_\pi^{(0,0)}=1$. We are not able to give a closed 
formula for the sum. However,
very useful information can be obtained by noticing that
\[
  \frac{A_{Q + \pi}^{(0,0)}}{A_{Q - \pi}^{(0,0)}} 
  = \prod_{j=1}^N 
    \frac{\sin^2 \frac{2 j \pi - Q}{2 L}}{\sin^{2} \frac{2 j \pi + Q}{2 L}}, 
\]
and in the thermodynamic limit,
\[
  \frac{A_{Q + 2\pi}^{(0,0)}A_{Q -2 \pi}^{(0,0)}}
  { \left(A_Q^{(0,0)}\right)^2} 
     =  \frac{(\pi^2-Q^2)^2}{(2 L \sin \pi n)^4}\left[ 1  
      - \frac{2 \pi}{L} \cot \pi n + O(L^{-2})\right] .
\]
Here the $Q$ is extended outside the Brillouin zone.
Now it is straightforward to get the size- and filling-dependence of 
$A_Q^{(0,0)}$: 
\begin{equation}
  A_Q^{(0,0)} = \frac{f(Q)}{(L \sin \pi n)^{\alpha_Q}}
    \left[ 1-\alpha_Q\frac{\pi}{2 L} \cot \pi n +O(L^{-2}) \right] ,
   \label{eq:a00}
\end{equation}
where
\begin{equation}
  \alpha_Q = \frac{1}{2} \left( \frac{Q}{\pi} \right)^2 - \frac{1}{2} .
\end{equation}
Eq.~(\ref{eq:a00}) is also valid for $B_Q^{(0,0)}$, apart from the 
sign in the $1/L$ correction.

The $f(Q)$ is an even function of $Q$, $f(\pi)=1$, and it satisfies the
second order recurrence equation
\[
  \frac{ f(Q+2\pi)f(Q-2\pi)}{f^2(Q)} =  \frac{(\pi^2-Q^2)^2}{16} ,
\]
which can be reduced to 
\[
  \frac{ f(Q+\pi)}{f(Q-\pi)} = 
     \frac{\Gamma^2(Q/2 \pi)}{\Gamma^2(-Q/2 \pi)} 
                       \pi^{2 Q/\pi} ,
\]
and it follows that $f(3\pi)$, $f(5\pi)$ etc. are zero.
In the interval from $Q=0$ to $\pi$ it can be approximated as
\[
 \ln f(Q) \approx 
    -0.3047 + 0.3248 \frac{Q^2}{\pi^2} - 0.0201 \frac{Q^4}{\pi^4}
\]
with accuracy $0.0001$. 
Furthermore $\ln f(0) = -0.304637$.

\subsection{Low energy behavior}

As we can see in Fig.~\ref{fig:support}, for low energies 
$A_Q(k,\omega)$ has so called towers of excitations centered at momenta
$k= (N+1)(Q+2 p \pi)/L $, where $p$ is an integer. The largest 
weights are for the peaks in the tower with
$p=0$, the  next with $p=-1$ (if $Q>0$) or $p=1$ (if $Q<0$), and so on.
The lowest excitation in tower $p$ corresponds to a set
of densely packed quantum numbers ${\cal I}_j$ shifted by $p$.
 From the definition of the momenta $k_j$, this is equivalent to 
imposing a twist of wave-vector $Q+2 p \pi$. Therefore we can introduce 
$\tilde Q= Q+2 p\pi$, where $\tilde Q$ is not restricted to be in the 
Brillouin zone, but for $p\neq 0$ it has values outside. We define 
$A_{\tilde Q}(k,\omega)$ to describe the $p$-th tower, so that 
$A_{Q}(k,\omega)$ has contributions from each of the towers:
$A_{Q}(k,\omega) = \sum_p A_{\tilde Q}(k,\omega)$. 

Furthermore, we enumerate the peaks in a given tower with 
indices $i$ and $i'$, so that the energy and momentum of the peaks are, 
from Eqs.~(\ref{eq:kIQ}), (\ref{eq:ELHB}) and (\ref{eq:PLHB}):
\
\begin{eqnarray}
  E_{i,i'} &=& E_{\rm GS}^{N} + \varepsilon_{c} 
        + \frac{\pi}{2L} u_c \left(1+\frac{{\tilde Q}^2}{\pi^2}\right) 
       +  \frac{2 \pi u_c}{L} (i+i')  ,
 \label{eq:Echargefs}\\
  P_{i,i'} &=& k_{\tilde Q}+\frac{\tilde Q}{L} +\frac{2 \pi }{L}(i-i') , 
 \label{eq:Pchargefs}
\end{eqnarray}  
where we have neglected the $O(1/L^2)$ finite-size corrections. Here 
$\varepsilon_{c}=-2t \cos \pi n$ is the `Fermi energy', 
$u_c = 2 t \sin{\pi n}$ is the `Fermi (charge) velocity' and
$k_{\tilde Q}=n \tilde Q$ is the `Fermi momentum' of spinless fermions 
representing the charges.
By $A_{\tilde Q}^{(i,i')}$ we denote the weight of the peaks, and for 
convenience, we also
introduce the relative weights 
$a_{\tilde Q}^{(i,i')}=A_{\tilde Q}^{(i,i')}/A_{\tilde Q}^{(0,0)}$. 
The weight of the first few 
lowest-lying peaks can be  calculated explicitly by
Eqs.~(\ref{eq:defg})-(\ref{eq:def2ph}), as they are given by a finite 
number of particle-hole excitations.
  The degeneracy of each peak grows with $i$ and $i'$. 
Here we assumed that the dispersion relation is linear near the Fermi 
level with velocity $u_c$. Clearly, this picture is valid for 
energies small compared to bandwidth.

  From Eq.~(\ref{eq:def1ph}) we get the relative weights 
$a_{\tilde Q}^{(i,i')}$, e.g. 
$a_{\tilde Q}^{(1,0)}$ is given as:
\[
  a_{\tilde Q}^{(1,0)} =
    \frac{
      \sin^2(\frac{\pi+{\tilde Q}}{2L})
      \sin^2(\frac{\pi N+\pi}{L})
    }{
      \sin^{2}(\frac{\pi}{L}) 
      \sin^{2}(\frac{2 \pi N+\pi+{\tilde Q}}{2L})
    } .
\]
Introducing $w_j = ({\tilde Q}/\pi +j )^2/4$, the relative weights in 
the thermodynamic limit simplify so that:
\begin{eqnarray}
  a_{\tilde Q}^{(0,0)} &=& 1 ,
   \nonumber\\
  a_{\tilde Q}^{(1,0)} &=& w_1 ,
   \nonumber\\
  a_{\tilde Q}^{(2,0)} &=& \frac{1}{2^2} w_1 (w_{-1} + w_3) ,
    \nonumber\\
  a_{\tilde Q}^{(1,1)} &=& w_{-1} w_1 ,
    \nonumber
\end{eqnarray}
and also  $a_{\tilde Q}^{(i,i')} = a_{-\tilde Q}^{(i',i)}$ holds. 
Note that some peaks are degenerate and therefore they are a sum of 
more contributions.
Now, it takes only one step to get the general formula
which reads (including the finite-size corrections):
\begin{eqnarray}
 a_{\tilde Q}^{(i,i')} &=& 
  \frac{ (1+\beta_{\tilde Q}) (2+\beta_{\tilde Q})
         \dots (i+\beta_{\tilde Q}) }{i!}
  \nonumber\\
   &&\times 
  \frac{ (1+\beta_{-{\tilde Q}}) (2+\beta_{-{\tilde Q}})
        \dots (i'+\beta_{-{\tilde Q}}) }{i'!} 
  \nonumber\\
  &&\times
  \left[      
      1 + \frac{(i+i')\pi-(i-i'){\tilde Q}}{L} \cot \pi n 
      +O(L^{-2}) 
  \right] ,
  \label{eq:aii}
\end{eqnarray}
where 
\begin{equation}
  \beta_{\pm {\tilde Q}} = 
  \left( \frac{1}{2} \pm \frac{{\tilde Q}}{2\pi} \right)^2 -1 .
\end{equation}
It can also be expressed with the help of the $\Gamma$-function, since 
\[
 \frac{(1\!+\!\beta_{\tilde Q})
       (2\!+\!\beta_{\tilde Q})\dots
       (i\!+\!\beta_{\tilde Q})}
      {i!} = 
  \frac{\Gamma(i\!+\!\beta_{\tilde Q}\!+\!1)}
       {\Gamma(i\!+\!1)\Gamma(\beta_{\tilde Q}\!+\!1)} .
\]
 The asymptotic expansion of the $\Gamma$-function gives
\begin{equation}
  \frac{\Gamma(i+\beta_{\tilde Q}+1)}{\Gamma(i+1)}
   \approx
    (i+1/2+\beta_{\tilde Q}/2)^{\beta_{\tilde Q}} ,
   \label{eq:Gasympt}
\end{equation}
which is a reasonable approximation apart from the $i=0$ peak.
Then, it follows that $a_{\tilde Q}^{(i,i')}$ has a power law behavior:
\begin{equation}
 a_{\tilde Q}^{(i,i')} = 
  \frac{(i+1/2+\beta_{\tilde Q}/2)^{\beta_{\tilde Q}} 
   (i'+1/2+\beta_{-{\tilde Q}}/2)^{\beta_{-{\tilde Q}}} }
   {\Gamma(\beta_{\tilde Q}+1)\Gamma(\beta_{-{\tilde Q}}+1)} . 
 \label{eq:aiipow}
\end{equation}
Note that the exponent $\alpha_{\tilde Q}$ in 
Eq.~(\ref{eq:a00}) is also given by
$\alpha_{\tilde Q} = \beta_{\tilde Q} + \beta_{-{\tilde Q}} + 1$.

We can clearly see the  manifestation of
the underlying conformal field theory: i) The finite-size corrections to
the energy and momentum [Eqs.~(\ref{eq:Echargefs}) 
and (\ref{eq:Pchargefs})] 
of the lowest lying peak in the tower determines the
exponents of the correlation functions; ii) The weights in the
towers are given by $\Gamma$-function.\cite{Cardy} 

The spectral function $A_{\tilde Q}(k,\omega)$ in the thermodynamic limit 
is given by
\begin{equation}
 A_{\tilde Q}(k,\omega) =  \sum_{i,i'} A_{\tilde Q}^{(i,i')}  
                   \delta( \omega \!-\! E_{i,i'} )
                   \delta_{ k, P_{i,i'}}  ,
  \label{eq:atl}
\end{equation}
and collecting everything together, Eqs.~(\ref{eq:a00}) and 
(\ref{eq:aiipow}-\ref{eq:atl}),  
for the low energy behavior of $A_Q(k,\omega)$ we get 
\begin{eqnarray}
 &&A_Q(k,\omega)  = \sum_p
  \frac{ f({\tilde Q}) \Theta(\omega-u_c |k-k_{\tilde Q}|) }
   {4 \pi u_c \sin(\pi n) 
     \Gamma(\beta_{{\tilde Q}} \!+\! 1)
     \Gamma(\beta_{-{\tilde Q}} \!+\! 1)} 
  \nonumber\\
  && \times
  \left[ 
    \frac{\omega \!-\! \varepsilon_c \!+\! u_c (k \!-\! k_{\tilde Q})}
         {4 \pi u_c \sin \pi n}
  \right]^{\beta_{\tilde Q}} 
  \left[
    \frac{\omega \!-\! \varepsilon_c \!-\! u_c (k \!-\! k_{\tilde Q})}
         {4 \pi u_c \sin(\pi n)}
  \right]^{\beta_{-{\tilde Q}}} .
  \label{eq:AQkom}
\end{eqnarray}
It is also worth mentioning the symmetry property 
$A_{Q}(k,\omega)=A_{-Q}(-k,\omega)$.
 The whole calculation can be repeated for the spectral function
$B_Q(k,\omega)$:
\begin{eqnarray}
 &&B_Q(k,\omega)  = \sum_p
  \frac{f({\tilde Q})\Theta(u_c |k+k_{\tilde Q}|-\omega)}
  {4 \pi u_c \sin(\pi n)
   \Gamma(\beta_{{\tilde Q}} \!+\! 1)
   \Gamma(\beta_{-{\tilde Q}} \!+\! 1)} \nonumber\\
  && \times
  \left[ 
    \frac{\varepsilon_c \!-\! \omega \!-\! u_c (k \!+\! k_{\tilde Q})}
         {4 \pi u_c \sin \pi n}
  \right]^{\beta_{-{\tilde Q}}} 
  \left[
    \frac{\varepsilon_c \!-\! \omega \!+\! u_c (k \!+\! k_{\tilde Q})}
         {4 \pi u_c \sin \pi n}
  \right]^{\beta_{{\tilde Q}}} .
\label{eq:BQkom}
\end{eqnarray}

We should note, however, that these expressions are restricted for the 
weights 
far from the edges of the towers, where the asymptotic expansion of the 
$\Gamma$-function, Eq.~(\ref{eq:Gasympt}), is valid. 
This is especially
true when $Q\rightarrow \pi$, where the correct result is 
$A_{\pi}(k,\omega)=\delta(\omega - \varepsilon_c - u_c [k - \pi n])$.
In other words, for the exponents close to $-1$ there can be a
considerable deviation from the power law behavior, and the spectral weight
accumulates along the edges of the towers. This behavior can be 
observed in Fig.~\ref{fig:AomkQ}, where the exponents 
are $\beta_{+Q}=-7/16$ and $\beta_{-Q}=-15/16$.

\subsubsection{Local spectral functions}

\begin{figure}
\epsfxsize=8.5 truecm 
\centerline{\epsffile{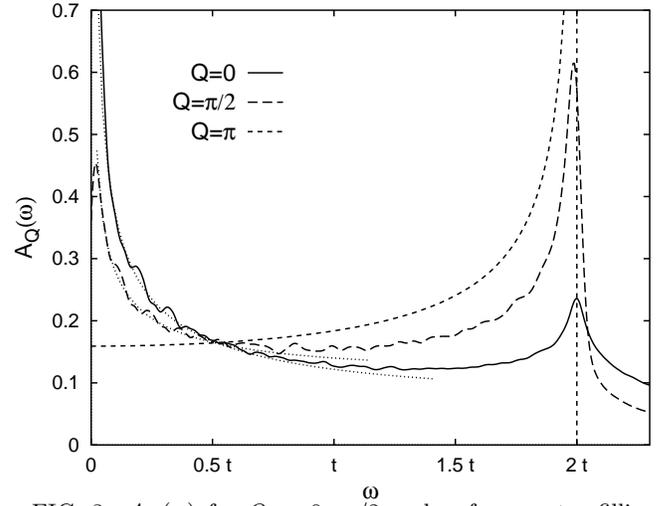}}
\caption{
  $A_Q(\omega)$ for $Q=0$, $\pi/2$ and $\pi$ for quarter filling ($L=300$,
$N=150$). For $Q=0$ the Van-Hove singularity is suppressed, and the weight
is mainly near the Fermi energy. $Q=\pi$ is equivalent to free-fermion 
case. The dotted line shows the low-energy approximation 
Eq.~(\protect{\ref{eq:aqapp}}).
}
\label{fig:AomQ}
\end{figure}

For the local ($k$-averaged) spectral function $A_{\tilde Q}(\omega)$
 the weight of the $j$-th peak, denoted by  $A_{\tilde Q}^{(j)}$, is 
\[
  A_{\tilde Q}^{(j)} = \frac{1}{L} \sum_{j'=0}^j A_{\tilde Q}^{(j',j-j')}
  .
\]
The  summation gives:
\begin{eqnarray}
   A_{\tilde Q}^{(j)} & = & \frac{1}{L}
     \frac{\Gamma(1+\alpha_{\tilde Q}+j)}
          {\Gamma(1+\alpha_{\tilde Q})\Gamma(1+j)} 
    A_{\tilde Q}^{(0,0)}
    \nonumber\\
    &&\times\left[ 
      1 + 
      j \frac{\pi}{L} \frac{\pi^2-{\tilde Q}^2}{\pi^2+{\tilde Q}^2} 
       \cot \pi n 
      +O(L^{-2}) 
    \right] .
  \nonumber
\end{eqnarray}
  If we put it together with Eqs.~(\ref{eq:a00}) and (\ref{eq:Echargefs}), 
and neglect the $1/L$ corrections, the local spectral function in the 
$L\rightarrow\infty$ limit reads:
\begin{equation}
    A_{Q}(\omega) \approx \sum_p \frac{1}{2 \pi u_c} 
    \frac{f({\tilde Q})}{\Gamma(\alpha_{ {\tilde Q}}+1)} 
    \left(\frac{\omega-\varepsilon_c}{2 \pi u_c \sin \pi n}
    \right)^{\alpha_{\tilde Q}} .
  \label{eq:aqapp} 
\end{equation}
For $B_{Q}(\omega)$ the $\omega-\varepsilon_c$ should be replaced by
$-\omega+\varepsilon_c$. We show $A_{Q}(\omega)$ for some
selected values of $Q$ in Fig.~\ref{fig:AomQ}.

\subsubsection{Momentum distribution function}

\begin{figure}
\epsfxsize=8.5 truecm 
\centerline{\epsffile{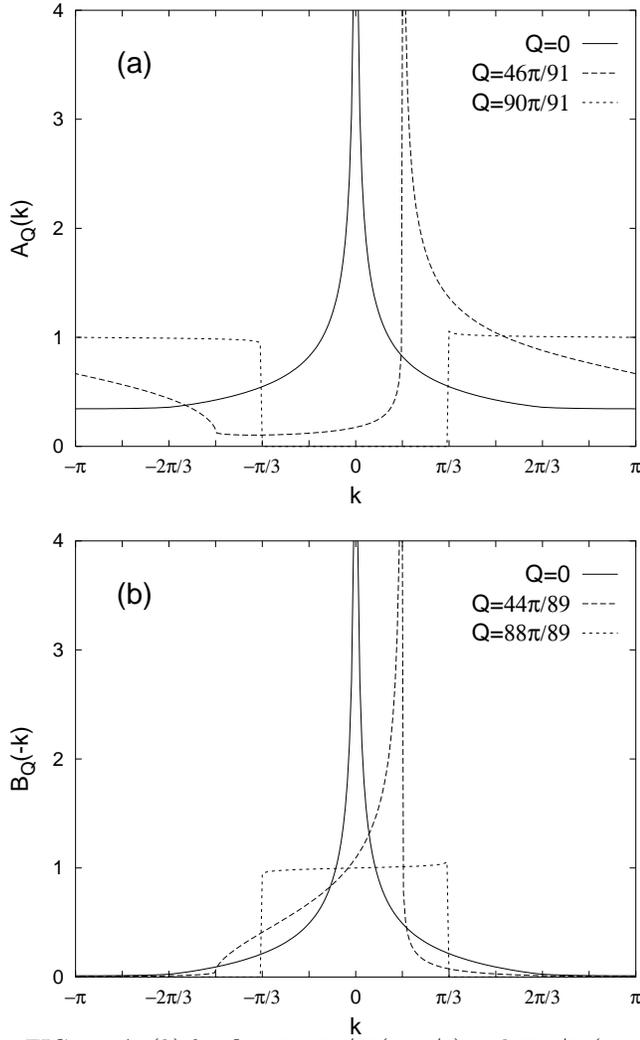}}
\caption{
  $A_Q(k)$ for $Q=0$, $46\pi/91$($\approx\pi/2$) 
  and $90\pi/91$($\approx\pi$)
(a) and $B_Q(-k)$ for $Q=0$, $44\pi/89$($\approx\pi/2$) and 
$88\pi/89$($\approx\pi$) (b) for $L=270$ and $N=90$. The evolution of the
weight and shape can be followed from the symmetric $Q=0$ case with 
the singularities at $k=0$ and $k=\pm 2\pi/3$ through the asymmetric 
$Q=\pi/2$ case with singularities at $k=\pi/6$ and $-\pi/2$ to the `normal'
distribution at $Q=\pi$. 
}
\label{fig:ABQ}
\end{figure}

Here we try to make some statements about $B_Q(k)$ in  Eq.~(\ref{eq:nkf}).
A na\"{\i}ve calculation in the low energy region is to sum up the 
weights near $k_{\tilde Q}$
\[
  B_{\tilde Q}^{(l)} = \sum_{i=0}^\infty
     \left\{ 
       \begin{array}{ll} 
            B_{\tilde Q}^{(l+i,i)},& \mbox{ \quad if $l\geq 0$ ;} \\
            B_{\tilde Q}^{(i,-l+i)},& \mbox{ \quad if $l < 0$ .} \\
       \end{array}
     \right. 
\]
Of course, one is aware that the summation includes high energies 
as well, where the equivalent for $b^{\tilde Q}_{i,i'}$ of
 Eq.~(\ref{eq:aii}) is not valid any more. However, the largest 
contributions come from the low energy regions and the error is not very 
large. 
We do not want to get precise values, but rather some qualitative results.
Neglecting the $O(1/L)$ corrections, the sum gives for $l \geq 0$:
\[
  B_{\tilde Q}^{(l)} \approx 
    \frac{\Gamma(-\alpha_{\tilde Q})\Gamma(1+l+\beta_{-{\tilde Q}})}
         {\Gamma(-\beta_{-{\tilde Q}})\Gamma(1+\beta_{-{\tilde Q}})
          \Gamma(l-\beta_{{\tilde Q}})} ,
\]
and for $l<0$ the $l$ and ${\tilde Q}$ should be 
replaced by $-l$ and $-{\tilde Q}$.
Again, we can use the asymptotic expansion of the $\Gamma$-function to get
\begin{equation}
  B_{\tilde Q}(k) \approx f({\tilde Q}) 
      \frac{\Gamma(-\alpha_{\tilde Q})}{\pi}
           \sin(-\pi \beta_{\pm {\tilde Q}}) 
            \left(
               \frac{|k-k_{\tilde Q}|}{2 \pi \sin \pi n}
            \right)^{\alpha_{\tilde Q}},
  \label{eq:BQk}
\end{equation}
where $\beta_{-{\tilde Q}}$ for  $k > k_{\tilde Q}$ and 
$\beta_{{\tilde Q}}$ for $k < k_{\tilde Q}$ should be taken in the 
argument of the sine.
It is interesting that, although the exponent of the singularity 
$\alpha_{\tilde Q}$ is the same for $k>k_{\tilde Q}$ and $k<k_{\tilde Q}$,
 there is a strong asymmetry due to 
the prefactor (a similar observation was made by Frahm and
Korepin\cite{frahm2}).
 In Fig.~\ref{fig:ABQ} this behavior is clearly observed.
For $Q\rightarrow\pi$ the correct result of 
$B_{\pi}(k) = \Theta(k_\pi-k) \Theta(k_\pi+k)$ is recovered.

\section{About the spin part}
\label{sec:spin}

 To calculate $C_{\sigma}(Q,\omega)$ and $D_{\sigma}(Q,\omega)$ given by 
Eqs.~(\ref{eq:defcomegaQ}), we need to know the 
energies and wave-functions of the spin part. They can be calculated
 from the usual spin-$\case{1}{2}$ Heisenberg Hamiltonian, see 
Eq.~(\ref{eq:Hseff}), taking  $N$ and $N\pm 1$ sites (spins).
 
For the $\tilde J \rightarrow 0$ case the excitation spectrum of the 
spins collapse, and then we can use the local, $\omega$ integrated 
functions $C_{\sigma}(Q)=\sum_\omega C_{\sigma}(Q,\omega)$ and 
$D_{\sigma}(Q)=\sum_\omega D_{\sigma}(Q,\omega)$. 
They are related to the spin transfer function 
$\omega_{j' \rightarrow j,\sigma}$, defined by Ogata and Shiba,\cite{shiba}
as was first noticed by Sorella and Parola.\cite{sorella2}
The spin transfer function gives the amplitude of removing a spin $\sigma$ 
at site $j'$ (here we choose $j'=0$) and inserting it at site $j$ ,
and can be given as
\[
  \omega_{0 \!\rightarrow \!j,\sigma} = 
  \langle \chi_N^{\rm GS} |
   \hat P_{j,j-1} 
   \dots 
   \hat P_{1,0} 
   \delta_{\sigma,S^z_0}
  | \chi_N^{\rm GS} \rangle ,
\]
where the operator $\hat P_{i,i+1}=2 {\bf S}_{i} {\bf S}_{i+1}
+\case{1}{2}$ permutes the spins at sites $i$ and $i+1$.
Then $C_{\sigma}(Q)$ and $D_{\sigma}(Q)$ read
\begin{eqnarray}
  C_{\sigma}(Q) &=& \frac{1}{N+1} 
     \Biggl[
       1 + 
       \sum_{j=0}^{N-1} e^{i (Q+\pi) (j+1)} \omega_{0\rightarrow j,\sigma} 
     \Biggr] ,
\nonumber \\
  D_{\sigma}(Q) &=& \frac{1}{N-1} \sum_{j=0}^{N-2} e^{i (Q+\pi) j} 
        \omega_{0\rightarrow j,\sigma} .
  \label{eq:CDw}
\end{eqnarray}
In particular, $\omega_{0 \rightarrow 0,\sigma}= N_\sigma/N$, and it 
follows that $ \sum_Q C_{\sigma}(Q) = 1$ 
and $\sum_Q D_{\sigma}(Q) = N_\sigma/N$.

 We are interested in these quantities for two particular cases: the
isotropic Heisenberg model because it is physically relevant, and 
the $XY$-model because it allows analytical calculations. 
We first  consider the $XY$-model 
because the simplicity of that case makes it more convenient to 
introduce the basic ideas.

\subsection{XY model}

In this special case the spin problem can be mapped to noninteracting 
spinless
fermions  using the Wigner-Jordan transformation. It means that the
eigenenergies and wave functions are known, and we can calculate 
$D_{\sigma}(Q,\omega)$ and $C_{\sigma}(Q,\omega)$ analytically. We are 
facing
a similar problem - the orthogonality catastrophe - as when we calculated
the $A_Q(\omega,k)$, but now it comes from  
the overlaps between states with different number of sites. For 
convenience,
we choose the spinless fermions to represent the 
$\bar\sigma$ spins, so that the operator $\hat Z^\dagger_{0,\sigma}$
($\hat Z^{\phantom{\dagger}}_{0,\sigma}$) only adds (removes)
a site and does not
change the number of fermions, which we fix to be $N_{\bar\sigma}$.
Then we have to evaluate matrix elements like
$\langle \tilde \chi_{N+1} |
 \hat Z^\dagger_{0,\sigma} | \tilde \chi_N^{\rm GS} \rangle $ 
and 
$\langle \tilde \chi_{N- 1} | 
 \hat Z^{\phantom{\dagger}}_{0,\sigma} | \tilde \chi_N^{\rm GS} \rangle $,
where in the $| \tilde \chi_N^{\rm GS} \rangle $ the $0$ site is 
unoccupied 
and the fermions are on sites $l=1...N$ and from site $l=1$ they hop
to $l=N$ skipping the $l=0$ site. For simplicity, we consider cases when 
the number of spin up and down fermions is odd ($N$ is even), 
so that we do not have to worry
about extra phases arising from the Jordan-Wigner transformation. Then the
momentum of the ground state $| \tilde \chi_N^{\rm GS} \rangle $ 
is $P_{\rm GS} = \pi$.
Let us denote by $k'$ the momenta of fermions on a $N\pm 1$ site lattice, 
quantized as 
$k'_j=2\pi {\cal J}'_j/(N\pm 1)$ and by 
$k$ the momenta of fermions on a $N$ site lattice, quantized as 
$k=2\pi {\cal J}_j/N$,
where ${\cal J}_j$ and ${\cal J}'_j$ are integers 
($j=1,\dots, N_{\bar\sigma}$), and by
 $f$ and $f^\dagger$ the operators 
of the spinless fermions. The energy and momentum of the state are: 
\begin{eqnarray}
  E &=& J_{\rm XY} \sum_{j=1}^{N_{\bar \sigma}} \cos k'_j ,
  \label{eq:EspinfsXY}\\
  P &=& \sum_{j=1}^{N_{\bar \sigma}}  k'_j .
  \label{eq:PspinfsXY}
\end{eqnarray}

To calculate the matrix element in $C_\sigma(Q,\omega)$, 
see Eq.~(\ref{eq:defcomegaQ}), we need the following anti-commutation
relation:
\begin{eqnarray}
    \bigl\{ f^\dagger_{k'}, f^{\phantom{\dagger}}_{k} \bigr\}
    &=& \frac{1}{\sqrt {N(N+1)}}\sum_{l=1}^N \sum_{l'=0}^N e^{ik'l'-ikl}  
    \bigl\{ f^\dagger_{l'}, f^{\phantom{\dagger}}_{l} \bigr\} \nonumber\\
    &=& \frac{1}{\sqrt {N(N+1)}} e^{i k /2} 
       \frac{\sin(k'/2)}{\sin([k-k']/2)} ,
  \nonumber
\end{eqnarray}
and the matrix element $ \bigl| 
    \langle \chi_{N+1} (\{{\cal J}'\}) |
 \hat Z^\dagger_{0,\sigma} | \chi_N^{\rm GS} \rangle
  \bigr|^2 =  |  \langle 0|
      f^{\phantom{\dagger}}_{k_{N_{\bar \sigma}}} 
      \dots 
      f^{\phantom{\dagger}}_{k_2}
      f^{\phantom{\dagger}}_{k_1}
      f^\dagger_{k'_1} 
      f^\dagger_{k'_2} 
      \dots 
      f^{\dagger}_{k'_{N_{\bar \sigma}}}
   |0\rangle |^2
$ is again given by a Cauchy determinant, which can be expressed as a 
product:
\begin{eqnarray}
 && [N(N+1)]^{-{N_{\bar\sigma}}} 
  \prod_{j=1}^{N_{\bar\sigma}} \sin^2 \frac{k'_j}{2} 
   \prod_{j>i} \sin^2 \frac{k_j-k_i}{2} 
  \nonumber \\
  &&\times
   \prod_{j>i} \sin^2 \frac{k'_j-k'_i}{2} 
   \prod_{i,j} \sin^{-2} \frac{k'_i-k_j}{2} .
   \label{eq:prodC}
\end{eqnarray}

Similarly, in the case of $D_\sigma(Q,\omega)$, the anticommutator is 
\[
    \bigl\{ f^\dagger_{k'}, f^{\phantom{\dagger}}_{k} \bigr\}
    = \frac{1}{\sqrt {N(N-1)}} e^{-i k' /2} 
       \frac{\sin(k/2)}{\sin([k'-k]/2)} ,
\]
 and the matrix element 
$\bigl| \langle \chi_{N-1} (\{{\cal J}'\}) |
  \hat Z^{\phantom{\dagger}}_{0,\sigma} | \chi_N^{\rm GS} \rangle
  \bigr|^2$  is equal to
\begin{eqnarray}
 && [N(N-1)]^{-{N_{\bar\sigma}}} 
  \prod_{j=1}^{N_{\bar\sigma}} \sin^2 \frac{k_j}{2} 
   \prod_{j>i} \sin^2 \frac{k_j-k_i}{2} 
  \nonumber \\
  &&\times
   \prod_{j>i} \sin^2 \frac{k'_j-k'_i}{2} 
   \prod_{i,j} \sin^{-2} \frac{k'_i-k_j}{2} .
   \label{eq:prodD}
\end{eqnarray}

  As soon as we have the product representation, it is straightforward to 
analyze the low energy behavior and also to obtain numerically 
$D(Q,\omega)$ and $C(Q,\omega)$ for larger system sizes.

\subsubsection{The low energy behavior}

The low energy spectra of $D_\sigma(Q,\omega)$ and $C_\sigma(Q,\omega)$
consist of towers centered at momenta  
$Q_{r,\sigma} =  2 r \pi \mu_{\sigma}$,
 where $r=1/2, 3/2,\dots$. 
 To analyze the low energy behavior in the tower labeled by $r$, 
we can proceed analogously
to the charge part: the weights in the tower of excitations,
$C^{(i,i')}_{r,\sigma} = c^{(i,i')}_{r,\sigma} C^{(0,0)}_{r,\sigma}$ and 
$D^{(i,i')}_{r,\sigma} = d^{(i,i')}_{r,\sigma} D^{(0,0)}_{r,\sigma}$, can 
be calculated from Eqs.~(\ref{eq:prodC}) and (\ref{eq:prodD}). 
The energy and momentum of the state $(i,i')$ can be calculated from 
Eqs.~(\ref{eq:EspinfsXY}) and (\ref{eq:PspinfsXY}) and neglecting
the $O(1/N^{2})$ corrections they read:
\begin{eqnarray}
  E_{i,i',r}^{(N\pm 1)}&=&  E_{\rm GS} 
    \pm \varepsilon_{\sigma} + 
     \frac{\pi}{N} u_\sigma
     \left(\gamma^{+}_{r,\sigma} + \gamma^{-}_{r,\sigma} + 2 \right) 
   \nonumber\\ && 
    + \frac{2\pi}{N} u_\sigma ( i+i') ,
 \label{eq:Espinfs}\\
  P_{i,i'}^{(N\pm 1)} &=& Q_{r,\sigma} \pm 
     \frac{ \pi}{N} 
     \left( \gamma^{+}_{r,\sigma} - \gamma^{-}_{r,\sigma} \right)
     +\frac{2 \pi}{N} (i-i') ,
   \label{eq:Pspinfs}
\end{eqnarray}
where
\begin{equation} 
 \gamma^{\pm}_{r,\sigma} = 
  \left(\frac{\mu_{\bar\sigma}}{2} \pm r \right)^2 -1 ,
  \label{eq:gammaXY}
\end{equation}
and the ``Fermi energy'' and the  velocity of the spins are:
\begin{eqnarray}
  \varepsilon_{\sigma} &=& J_{\rm XY} 
  \left( \mu_{\bar\sigma} \cos \pi \mu_{\bar\sigma}
       - \frac{1}{\pi} \sin \pi \mu_{\bar\sigma} \right) ,
   \nonumber\\
  u_{\sigma} &=& J_{\rm XY} \sin \pi \mu_{\bar\sigma} ,
\end{eqnarray}
and $\mu_{\bar\sigma}=N_{\bar\sigma}/N$.
 
The relative weights $d^{(i,i')}_{r,\sigma}$ can be calculated from 
Eq.~(\ref{eq:prodD}), e.g.:
\begin{eqnarray}
  d^{(0,1)}_{1/2,\sigma}
      &=& \frac{ \sin^2\frac{\pi(1+N+N_{\bar\sigma})}{N-1}}
               { \sin^2\frac{\pi}{N-1}}
          \frac{ \sin^2\frac{\pi(1+N+N_{\bar\sigma})}{2(N^2-N)}}
               { \sin^2\frac{\pi(1+N-N_{\bar\sigma}+2N N_{\bar\sigma} )}
                     {2(N^2-N)}}  
\nonumber\\
&&\times
               \frac{ R^2 \left((N+N_{\bar\sigma}+1)/2\right)}
	           { R^2 \left((N+N_{\bar\sigma}-1)/2\right)} ,
\nonumber
\end{eqnarray}  
where
\[
  R(l) = \prod_{j=0}^{N_{\bar\sigma}-1} 
    \sin \left(\frac{\pi l}{N(N-1)}+\frac{\pi j}{N}\right) ,
\]
and the other $d^{(i,i')}_{r,\sigma}$ are similar. 
In the thermodynamic limit, $N \rightarrow \infty$,
the weight $d^{(0,1)}_{1/2,\sigma}$ simplifies to
\begin{equation}
 d^{(0,1)}_{1/2,\sigma} 
    = \left(\frac{1}{2}+\frac{\mu_{\bar\sigma}}{2}\right)^2
                    \left[1+O(\ln L/L)\right] .
  \label{eq:dgamma10}
\end{equation}
Neglecting the finite-size corrections, for general $(i,i')$ 
and $r$ we get:
\begin{equation}
 d^{(i,i')}_{r,\sigma} = \frac{\Gamma(i+\gamma_{r,\sigma}^{-}+1)}
                {\Gamma(\gamma_{r,\sigma}^{-}+1)\Gamma(i+1)} 
             \frac{\Gamma(i'+\gamma_{r,\sigma}^{+}+1)}
                {\Gamma(\gamma_{r,\sigma}^{+}+1)\Gamma(i'+1)} ,
 \label{eq:dgamma}
\end{equation}
where the exponents $\gamma_{r,\sigma}^{\pm}$ are defined in 
Eq.~(\ref{eq:gammaXY}) and the weights again follows the prescription 
of the conformal theory, with strong logarithmic 
finite-size corrections however.
 A similar analysis can be done for 
$C_\sigma(Q,\omega)$. From the above and Eq.~(\ref{eq:defcomegaQ}) 
we obtain
\begin{eqnarray}
  D_\sigma(Q,\omega) &\approx& \sum_r g(r,\mu_\sigma)
   \left[\varepsilon_{\sigma}-\omega + u_\sigma (Q-Q_{r,\sigma})
   \right]^{\gamma_r^-}
   \nonumber\\
   &&\times
   \left[\varepsilon_{\sigma}-\omega  - u_\sigma (Q-Q_{r,\sigma})
   \right]^{\gamma_r^+} 
    \nonumber\\
   &&\times
  \Theta(\varepsilon_{\sigma}-\omega + u_\sigma |Q-Q_{r,\sigma}|)
  \label{eq:DQom}
\end{eqnarray}
and
\begin{eqnarray}
  C_\sigma(Q,\omega) &\approx& \sum_r g(r,\mu_\sigma)
   \left[\omega-\varepsilon_{\sigma} + u_\sigma (Q-Q_{r,\sigma})
   \right]^{\gamma_r^+}
   \nonumber\\
   &&\times
   \left[\omega-\varepsilon_{\sigma} - u_\sigma (Q-Q_{r,\sigma})
   \right]^{\gamma_r^-}
   \nonumber\\
   &&\times
   \Theta(\omega-\varepsilon_{\sigma} - u_\sigma |Q-Q_{r,\sigma}|)
  \label{eq:CQom}
\end{eqnarray}
where $g(r,\mu_\sigma)$ are numbers which can be determined numerically.

 We immediately see that the $C_\sigma(Q)$ and $D_\sigma(Q)$ are 
singular at $Q=Q_{r,\sigma}$:
\[
  C_\sigma(Q), D_\sigma(Q) \propto
  |Q-Q_{r,\sigma}|^{\eta_{r,\sigma}}
\]
 with exponent 
\[
 \eta_{r,\sigma} = \gamma^{+}_{r,\sigma} + \gamma^{-}_{r,\sigma} +1
\]
and they are strongly asymmetric around $Q_{r,\sigma}$, 
as we can conclude from the analog of Eq.~(\ref{eq:BQk}).

For the non-magnetic case ($\mu_{\sigma}=\mu_{\bar \sigma}=1/2$), the 
singularity is at $Q_r=\pi/2$ for all the towers, and the exponents of
 the main singularity ($r=1/2$) are 
$\gamma_{1/2}^- =-15/16$ and $\gamma_{1/2}^+ =-7/16$, furthermore 
$\eta_{1/2} = -3/8$.
 
\subsection{Heisenberg model}

\begin{figure}[t]
\epsfxsize=8.5 truecm 
\centerline{\epsffile{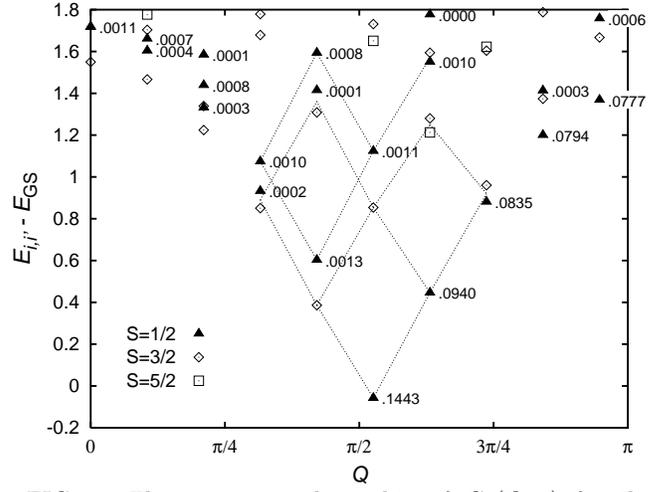}}
\caption{The support and weights of
$C_\sigma(Q,\omega)$ for the $N=18$ spin Heisenberg model.
 The symbols represent the 
excitations of the final states (19 spins), where the total spin is also 
indicated. The numbers near solid triangles give the
weight of that particular state. Due to selection rule the matrix elements
are zero with higher spin states denoted by open symbols.
The dotted lines are a guide to the eyes and show the  $r=1/2$ and 
$r=-3/2$ towers. 
}
\label{fig:weights}
\end{figure}

Although the Heisenberg model is solvable by Bethe-ansatz and in 
principle the
wave functions are known, it is too involved to give the matrix elements 
of $C_{\sigma}(Q,\omega)$ and $D_{\sigma}(Q,\omega)$. The simplest 
alternative way is 
 exact diagonalization of small clusters and DMRG\cite{white} extended to
dynamical properties.\cite{karendyn} 
We have used both methods 
to calculate the weights for system sizes up to $N=24$ and $N=42$, 
respectively.
A typical distribution of the weights for $C_{\sigma}(Q,\omega)$ for 
zero magnetization 
is given in  Fig.~\ref{fig:weights}. There are several features to be 
observed: i) Due to selection rules, the nonzero matrix elements are with 
the $S=1/2$ final states only;
ii) The weight is concentrated along the lower edge
of the excitation spectra in the interval 
 $\pi/2 \geq Q \geq \pi$; iii) There are two, almost overlapping  towers 
visible corresponding to $r=1/2$ and $r=-3/2$. Our interpretation of the
spectrum is that the weight mostly follows the dispersion of the 
spinon of Faddeev and Takhtajan,\cite{spinon} since the final states 
have an odd number of spins,
 thus there can be a single spinon in the spectrum and it
has a cosine-like dispersion. It is also surprising that for 
$C_{\sigma}(Q,\omega)$ more than 97\% and for $D_{\sigma}(Q,\omega)$ more
than 99\% of the total weight is found in this spinon branch. This 
behavior is similar to that discussed by Talstra, Strong and 
Anderson,\cite{talstra}
where they added two spins to the spin wave function.

We can also try to analyze the low energy behavior from the conformal field
theory point of view. Namely, from the Bethe-ansatz solutions the 
finite-size corrections to the energy are 
known\cite{woynaPRL,woyna87,alcaraz} 
and they are also given by Eqs.~(\ref{eq:Espinfs}) 
and (\ref{eq:Pspinfs}) apart from $\ln(N)/N$ corrections, with 
\begin{equation}
 \gamma^{\pm}_{r,\sigma} = 
  \left(\frac{\mu_{\bar\sigma}}{2 \xi} \pm \xi r \right)^2-1 .
  \label{eq:gamma}
\end{equation} 
 For zero magnetization the velocity $u_\sigma$ reads $ \pi \tilde J / 2$, 
the energy is  $\varepsilon_{\sigma}=-\tilde J \ln 2$ and  
$\xi=1/\sqrt{2}$, and the exponents are $\gamma_{1/2}^- =-1$ and 
$\gamma_{1/2}^+ =-1/2$, very close to the $XY$ exponents 
($-15/16$ and $-7/16$, respectively). 
For arbitrary magnetization $u_\sigma$, $\varepsilon_{\sigma}$ and $\xi$ 
are to be calculated from integral equations.\cite{woyna87}

\begin{figure}[b]
\epsfxsize=8.5 truecm 
\centerline{\epsffile{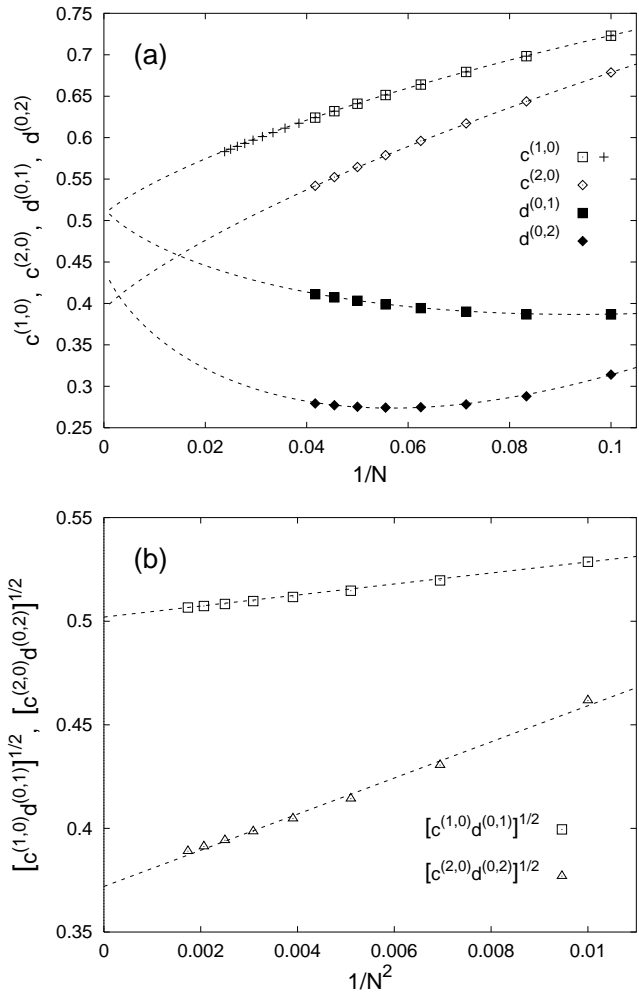}}
\caption{
 The relative weights $c^{(1,0)}$, $c^{(2,0)}$, $d^{(0,1)}$ and $d^{(0,2)}$
as a function of the system size calculated by exact diagonalization 
(squares and triangles) and by DMRG (crosses) for the $r=1/2$ tower. 
The dashed line represents a fit to 
$a_0+a_1/N+a_2 \log(N)/N$ form and it is reasonably close to the 
theoretical values $0.5$ and $0.375$ in the thermodynamic limit (a). 
The opposite sign of logarithmic corrections cancels if we make the 
products $[c^{(1,0)}d^{(0,1)}]^{1/2}$ and $[c^{(2,0)}d^{(0,2)}]^{1/2}$ (b).
}
\label{fig:cts}
\end{figure}

 Also, we check if Eq.~(\ref{eq:dgamma}) is satisfied for the $r=1/2$ tower
in Fig.~\ref{fig:cts}. Namely, it tells us that $c^{(1,0)}=d^{(0,1)}=1/2$
and $c^{(2,0)}=d^{(0,2)}=3/8$, apart from finite-size corrections which we
assumed to be of the same form as in the case of the $XY$ model in 
Eq.~(\ref{eq:dgamma10}). We believe that this method can also be used
to determine exponents in a more general cases as well.

Another interesting point is that the exponent
 $\gamma_{1/2}^- =-1$ already indicates that $c^{(0,1)}$ vanishes, in
 agreement with the selection rules. However, there is still some weight 
for $c^{(0,2)}$, which comes from $S=1/2$ bound
states of spinons. We do not know the finite-size scaling of 
that weight, i.e. if it disappears in the thermodynamic limit or not.

  Now, if we recall that $D_\sigma(Q) = \sum_\omega D_\sigma(Q,\omega)$, 
then it follows (see Eq.~(\ref{eq:BQk})) that the contribution to $D_Q$ 
for $Q>\pi/2$ is
strongly suppressed, and we see essentially the contributions from the 
$r=3/2$ tower.
 Since the contribution to $C(Q,\omega)$ and $D(Q,\omega)$ come 
mostly 
from the lower edge of excitation spectrum, we can use the 
approximations 
\begin{eqnarray}
  C_\sigma(Q,\omega) &=& C_\sigma(Q) 
    \delta(\omega - \varepsilon_s - \varepsilon_Q) ,
    \nonumber\\
  D_\sigma(Q,\omega) &=& D_\sigma(Q) 
   \delta(\omega - \varepsilon_s + \varepsilon_Q) ,
  \nonumber
\end{eqnarray}
where $\varepsilon_Q$ is the des Cloizeaux-Pearson 
dispersion\cite{desCLoizeaux}
\[
 \varepsilon_Q = \frac{\pi}{2}  \tilde J |\sin( Q-\pi/2 )| .
\]

The $C_\sigma(Q)$ and $D_\sigma(Q)$ can be calculated 
numerically for small clusters (typically up to $N=26$ with exact
diagonalization and $N=70$ with DMRG) for the non-magnetic case 
(see Refs.~\onlinecite{sorella2,local}).
The $(N+1)  C_{\sigma}(Q)$ and $(N-1)  D_{\sigma}(Q)$ seems to have 
small finite-size effect, as follows from Eq.~(\ref{eq:CDw}), and the
singularity in the non-magnetic case is given by $\eta_{1/2}=-1/2$, 
as already noticed by Sorella and Parola.\cite{sorella2}

\begin{figure}
\epsfxsize=8.5 truecm  
\centerline{\epsffile{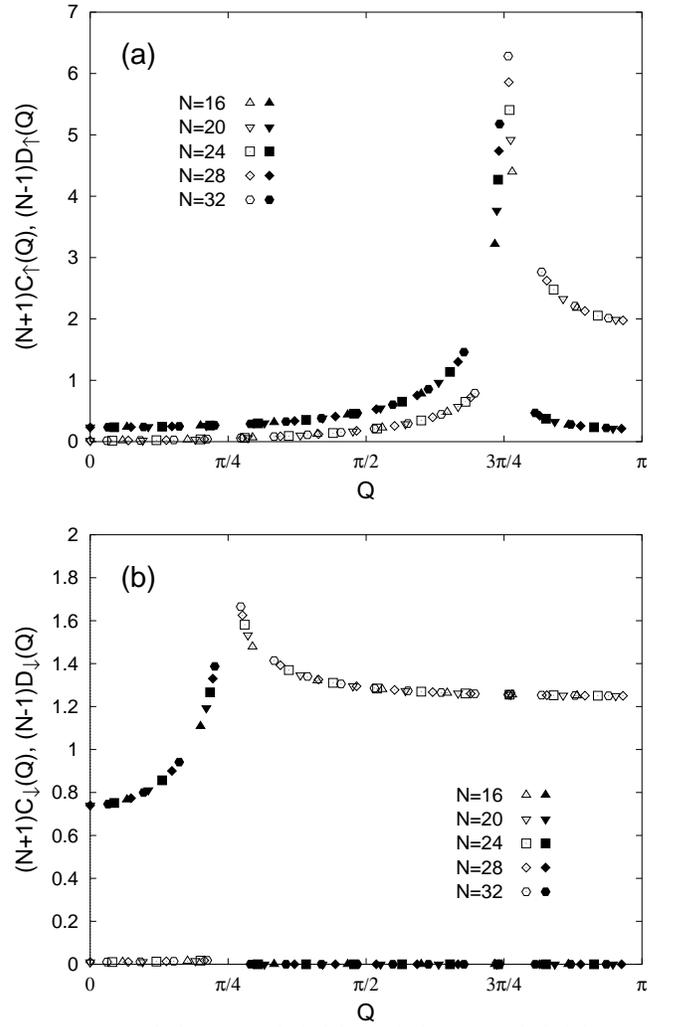}}
\caption{
 $C_\uparrow(Q)$ and $D_\uparrow(Q)$ (a), $C_\downarrow(Q)$ and 
$D_\downarrow(Q)$  (b) for finite magnetization
$N_\uparrow/N=3/4$ with singularity at $Q=3\pi/4$ and $Q=\pi/4$, 
respectively. 
The solid symbols stands for $D_{\sigma}(Q)$ and open for $C_{\sigma}(Q)$.
}
\label{fig:QM}
\end{figure}

  We have also calculated $C_\sigma(Q)$ and $D_\sigma(Q)$ for the system 
with finite magnetization $N_\downarrow/N=1/4$ (see Fig.~\ref{fig:QM}). 
There $Q_\uparrow = 3\pi/4$, $Q_\downarrow = \pi/4$
and the exponents are  $\eta_{1/2,\uparrow} = -0.58 \pm 0.03$ 
and $\eta_{1/2,\downarrow} = -0.25 \pm 0.03$. These exponents are 
consistent with $\xi= 0.87 \pm 0.02$ and in surprisingly good
agreement with the simple formula given by Frahm and Korepin\cite{frahm2}
$\xi \approx 1-\mu_\downarrow/2$ valid 
in a large magnetic field.

\section{ The Green's function and the comparison with the Conformal Field 
Theory}
\label{sec:CFT}

The real space Green's function can be calculated from the 
spectral functions as 
\[
  G(x,t) = \int_{-\pi}^{\pi} d k \int_{-\infty}^{\infty}d \omega 
   e^{i\omega t - i k x} 
   A(k,\omega) 
\]
for $t>0$ and $A(k,\omega)$ should be replaced by $B(k,\omega)$ for $t<0$, 
as follows from Eq.~(\ref{eq:GAB}). Then, from Eqs.~(\ref{eq:alhbca}), 
(\ref{eq:AQkom}) and (\ref{eq:CQom}) it follows that:
\begin{eqnarray}
G(x,t>0) &\approx& \sum_{p,r} 
  \frac{ c_{p,r} e^{-i \tilde Q_r x N/L}}
       {(x-u_c t)^{\beta_{\tilde Q_r}+1}(x+u_c t)^{\beta_{-\tilde Q_r}+1}}
    \nonumber\\
    &&\times 
    \frac{1}{(x-u_s t)^{\gamma^+_r +1} (x+u_c t)^{\gamma^-_r +1}} ,
\end{eqnarray}
where $\tilde Q_r$ was defined as $Q_r+2\pi p$, 
furthermore $c_{p,r}$ are numbers.
The charge velocity $u_c$ is the same one as in Eq.~(\ref{eq:Echargefs}), 
while the spin velocity is $u_s = u_\sigma / n$, where $u_\sigma$ was 
defined in Eq.~(\ref{eq:Espinfs}).
The Green's function has singularities at different momenta, depending 
on the actual quantum numbers $p$ and $r$, see Table.~\ref{tab:sing} for 
details.

On the other hand, according to the conformal field 
theory,\cite{frahm,kawakami} a correlation functions 
$\langle \phi(x,t) \phi(0,0) \rangle$ reads:
\[
 \sum_{D_c,D_s} 
   \frac{c_{D_c,D_s}
         e^{-2 i [D_c k_\uparrow + (D_c+D_s) k_\downarrow] x}}
      {(x\!-\!u_c t)^{2 \Delta^+_c} (x\!+\!u_c t)^{2 \Delta^-_c}  
       (x\!-\!u_s t)^{2 \Delta^+_s} (x\!+\!u_s t)^{2 \Delta^-_s}} 
     ,
\]
where the exponents
\begin{eqnarray}
  2\Delta^{\pm}_c = \left( Z_{cc} D_c + Z_{sc} D_s \pm
      \frac{Z_{ss} \Delta N_c - Z_{cs} \Delta N_s}{2 \det Z}  \right)^2 ,
  \nonumber\\
  2\Delta^{\pm}_s = \left( Z_{cs} D_c + Z_{ss} D_s \pm
      \frac{Z_{cc} \Delta N_s - Z_{sc} \Delta N_c}{2 \det Z}  \right)^2 ,
\end{eqnarray}
 are related to the finite-size corrections:
\begin{eqnarray}
 E-E_0 &=& \frac{2 \pi}{N} u_c \left(\Delta^+_c + \Delta^-_c \right)
         + \frac{2 \pi}{N} u_s \left(\Delta^+_s + \Delta^-_s \right),
      \label{eq:EfsBA}\\
P-P_0 &=& 2 D_c k_\uparrow + 2(D_c+D_s) k_\downarrow \nonumber \\
     && +  \frac{2 \pi}{N} 
     (\Delta^+_c - \Delta^-_c + \Delta^+_s - \Delta^-_s) . 
      \label{eq:PfsBA}
\end{eqnarray}
and $c_{D_c,D_s}$ are numbers.
The quantum numbers  $D_c$, $D_s$, $\Delta N_c$ and $\Delta N_s$ 
characterize the excitations and are related to $p$ and $r$ as 
given in Table.~\ref{tab:qn}. 
The $Z$'s are the elements of the so called dressed charge matrix. 
It can be calculated from Bethe
Ansatz solution of the Hubbard model, and in the large $U$ limit they read:
\[
  \begin{array} {ll} 
   Z_{cc} =  1 & Z_{cs}=0 \\
   Z_{sc} = \mu_\downarrow & Z_{ss} = \xi  ,\\
  \end{array}
\]
where $\xi$ can be obtained solving an integral equation. For the
non-magnetic case $\mu_\downarrow=1/2$ and $\xi=1/\sqrt{2}$.

Then we are ready to identify the exponents:
$\beta_{\pm\tilde Q_r}+1=2\Delta^\pm_c$ and 
$\gamma^\pm_{r}+1=2\Delta^\pm_s$, and
in this way we can directly see the validity of the CFT in the 
large-$U$ limit.

In case of the $t-J_{XY}$ model no Bethe Ansatz result is known, but 
using the analogy with the isotropic case, the exponents are 
readily obtained using the substitution
$Z_{cc} \rightarrow  1$, $Z_{cs} \rightarrow 0$, 
$Z_{sc} \rightarrow \mu_{\downarrow}$ and $Z_{ss} \rightarrow 1$.

\section{Conclusions}
\label{sec:conclu}

  To conclude, we have shown that for some special cases 
  the spectral functions of the 1D Hubbard can be calculated using the 
  spin-charge factorized wave-function, which implies that the spectral
  functions are given as a convolution involving the charge and spin parts.
  Analytical calculations are possible for the charge part and for the 
  spin part
  in the case of the $XY$ model. The low energy behavior turns out
  to be fully consistent with the predictions of the conformal field 
  theory,
  i.e. the exponents are given by the finite-size corrections to the energy
  and momentum, and the weights are given by $\Gamma$-function. Based on
  this, we propose a new way to determine the exponents of the correlation
  functions. Furthermore, we argue that when the exponents of the 
  correlation functions are close to integers, the  
  Luttinger-liquid power-law behavior of the correlation 
  functions should be taken with care, as it comes
  from the asymptotic expansion of the $\Gamma$ function.


\begin{table}
  \caption{Sum rule, Eq.~(\protect{\ref{eq:srAB}}), for $Q=0$ including 
one, two and three particle-hole  excitations, $N=L/2$.}
  \label{tab:sr}
  \begin{tabular}{rccc}
  L   &        1 p-h   &      1+2 p-h   &    1+2+3 p-h \\
\hline
   4 &   0.50000000  &  0.50000000  &  0.50000000 \\
  12 &   0.46477280  &  0.49989083  &  0.49999999 \\
  20 &   0.43436168  &  0.49933463  &  0.49999968 \\
  28 &   0.41165708  &  0.49844924  &  0.49999808 \\
  36 &   0.39388871  &  0.49738700  &  0.49999428 \\
  44 &   0.37941227  &  0.49623473  &  0.49998778 \\
  52 &   0.36725942  &  0.49504054  &  0.49997842 \\
  60 &   0.35682437  &  0.49383182  &  0.49996622 \\
\end{tabular}
\end{table}

\begin{table}
  \caption{The momenta for which the Greens function $G_\sigma(x,t>0)$ 
is singular.}
  \label{tab:sing}
  \begin{tabular}{ccccc}
  $r$ & $p=-1$ & $p=0$ & $p=1$ \\
  \hline
  -3/2  & \dots & $-3 k_\sigma$ & $-k_\sigma + 2 k_{\bar\sigma}$  \\
  -1/2  & \dots & $-k_\sigma$ & $k_\sigma + 2 k_{\bar\sigma}$  \\
   1/2  & $-k_\sigma - 2 k_{\bar\sigma}$ & $k_\sigma$ & \dots \\
   3/2  & $k_\sigma - 2 k_{\bar\sigma}$ & $ 3 k_\sigma$ & \dots  \\
  \end{tabular}
\end{table}

\begin{table}
  \caption{The correspondence between the Bethe Ansatz quantum numbers and 
$p$ and $r$}
  \label{tab:qn}
  \begin{tabular}{cccrr}
   $\sigma$     & $D_c$  & $D_s$   & $\Delta N_c$ & $\Delta N_s$ \\
   \hline
   $\uparrow$   & $p+r$  & $-r$    & $1$      &   $0$  \\
   $\downarrow$ & $p$    & $r$     & $1$      &   $1$  \\
   $\uparrow$   & $-p-r$  & $r$    & $-1$      &   $0$  \\
   $\downarrow$ & $-p$    & $-r$   & $-1$      &   $-1$  \\
  \end{tabular}
\end{table}
\newpage

\newpage

\end{multicols}
\end{document}